\documentclass{article}
\usepackage{graphicx} 
\usepackage[left=1in,right=1in,top=1in,bottom=1in]{geometry}
\usepackage{amsmath}
\usepackage{amsthm,amssymb,bm,xcolor}
\usepackage{authblk}
\usepackage{mathtools}
\usepackage{fourier}
\usepackage{bbm}
\usepackage[colorlinks=true,citecolor=blue]{hyperref}
\numberwithin{equation}{section}

\title{Quasi-continuum descriptions of rarefaction and dispersive shock waves in Fermi-Pasta-Ulam lattices with Hertzian potentials}
\author[1]{Su Yang\thanks{Corresponding author. Email: suyang@umass.edu}}
\affil[1]{Department of Mathematics and Statistics, University of Massachusetts Amherst, Amherst, MA 01003-4515, USA}
\date{\small\today}

\begin{document}

\maketitle

\begin{abstract}
 
In the present work, we review two well-established quasi-continuum models of a Fermi-Pasta-Ulam lattice with Hertzian type potentials, and utilize these two models to approximate the discrete dispersive shock waves (DDSWs) which are numerically observed in the simulation of the lattice. To perform analysis on the various characteristics of the DDSW, we analytically derive the Whitham modulation equations of the two quasi-continuum models, which govern the slowly varying spatial and temporal dynamics of distinct parameters of the periodic solutions. We then perform a very useful reduction of the Whitham modulation system to gain a system of initial-value problems whose solutions can provide important insights on edge features of the DSWs such as their edge speeds. In addition, we also study the numerical rarefaction waves (RWs) of the lattice based on the two quasi-continuum models. In particular, we analytically compute and compare their self-similar solutions with the numerical discrete RW of the lattice. These comparisons made for both DSWs and RWs reveal to be reasonably good, which suggest the impressive performance of both quasi-continuum models.
    
\end{abstract}


\tableofcontents

\section{Introduction}

Nonlinear waves are ubiquitous in various physics settings. Among these various nonlinear wave phenomena, the so-called dispersive shock wave \cite{Hoefer:2009} is defined as a non-stationary disperive wave structure which has raised considerably amount of interests recently. In particular, this wave structure is studied and numerically observed in lots of mathematical physics models including the well-known Toda lattice \cite{BIONDINI2024134315, toda2012theory, CHONG2024103352, PhysRevE.90.022905}, the nonlinear Schr\"odinger models \cite{mohapatra2025dambreaksdiscretenonlinear, yang2025dispersiveshockwavesperiodic,PhysRevA.110.023304,PhysRevLett.100.084504}, and some discrete lattices \cite{Yang_Biondini_Chong_Kevrekidis_2025, yang2025quasicontinuumapproximationsnonlineardispersive,https://doi.org/10.1111/sapm.12767, yang2025firstordercontinuummodelsnonlinear}. Moreover, besides its numerical emergence, the DSW can also be discovered in physics experimental environments \cite{PhysRevE.80.056602, PhysRevE.75.021304, PhysRevLett.120.194101}. Importantly, the core of the DSW is the periodic solutions associated with the model of interest, and the analysis of DSW needs to be performed based on the so-called Whitham modulation theory \cite{whitham2011linear,Abeya_2023,https://doi.org/10.1111/sapm.12651,Ablowitz_2018,PhysRevE.96.032225}, which reveals the evolution dynamics of the slowly-varying parameters of the periodic solutions. The DSW-fitting method \cite{EL201611} leads to a reduction of the Whitham modulation system, which then yields a system of simple-wave ordinary differential equations which encode the important information on multiple edge characteristics of the DSWs. The rarefaction waves, on the other hand, are classified as a simple wave \cite{whitham2011linear, 10.1093/oso/9780192843234.001.0001}. For this specific wave structure, dispersive models are also prototypical media for its numerical emergence (e.g. the standard  Korteweg-De Vries (KdV) equation \cite{EL201611}). Moreover, RWs can be analytically described by the self-similar solutions of the associated dispersionless version of the dispersive models, and this therefore provides a methodology for the analysis of the numerical RWs.

In the present work, we shall study both the DSW and RW observed in a FPU-type lattice. Previously, Ref.~\cite{CHONG2024103352} has proposed two integrable quasi-continuum models which are KdV and Toda equations to approximate only the DSW numerically observed in a granular chain. Instead, other works including \cite{Yang_Biondini_Chong_Kevrekidis_2025, yang2025firstordercontinuummodelsnonlinear} provide bi-directional and non-integrable uni-directional quasi-continuum models to approximate discrete DSWs in the granular crystal lattice. The present work shall analogously apply quasi-continuum approximations for the discrete nonlinear dispersive waves in the lattice, but however via two "non-standard' models: The so-called log-KdV equation and generalized-KdV equation with a H\"older continuous nonlinearity. 

This paper is structured as follows. In section \ref{sec: model intro}, we give a detailed description of the discrete lattice model which shall be the main focus of this work and simultaneously review two previously proposed quasi-continuum models. We next also review the important traveling solitary-wave solutions of both quasi-continuum models in section \ref{sec: solitary waves}. In section \ref{sec: periodic solutions}, we show the existence of the periodic solutions of the two quasi-continuum models via performing a phase-plane analysis. Following the periodic solutions, we list some important conservation laws and also discuss the Lagrangian structures of the quasi-continuum models in section \ref{sec: conservation laws} which serve as one necessary prerequisite of conducting Whitham analysis. In sections \ref{sec: Whitham theory} and \ref{sec: Whitham reductions}, we derive the relevant Whitham modulation equations via the method of averaging the conservation laws and perform a very useful reduction of such equations near both the harmonic and soliton limits, which yields a system of ODEs equipped with initial conditions. Next in section \ref{sec: RWs}, we analytically compute the self-similar solutions of the quasi-continuum models, which serve as approximations of the numerical discrete rarefaction waves of the lattice. We solve the two IVPs derived in section \ref{sec: Whitham reductions} to gain insights on distinct edge features of the DSWs, and compare these theoretical predictions with there numerically estimated counterparts in section \ref{sec: Numerical validation}. Finally, this paper ends with some open questions and future research directions as described in the conclusion section \ref{sec: conclusions}.

\begin{figure}[b!]
    \centering
    \includegraphics[width=0.8\linewidth]{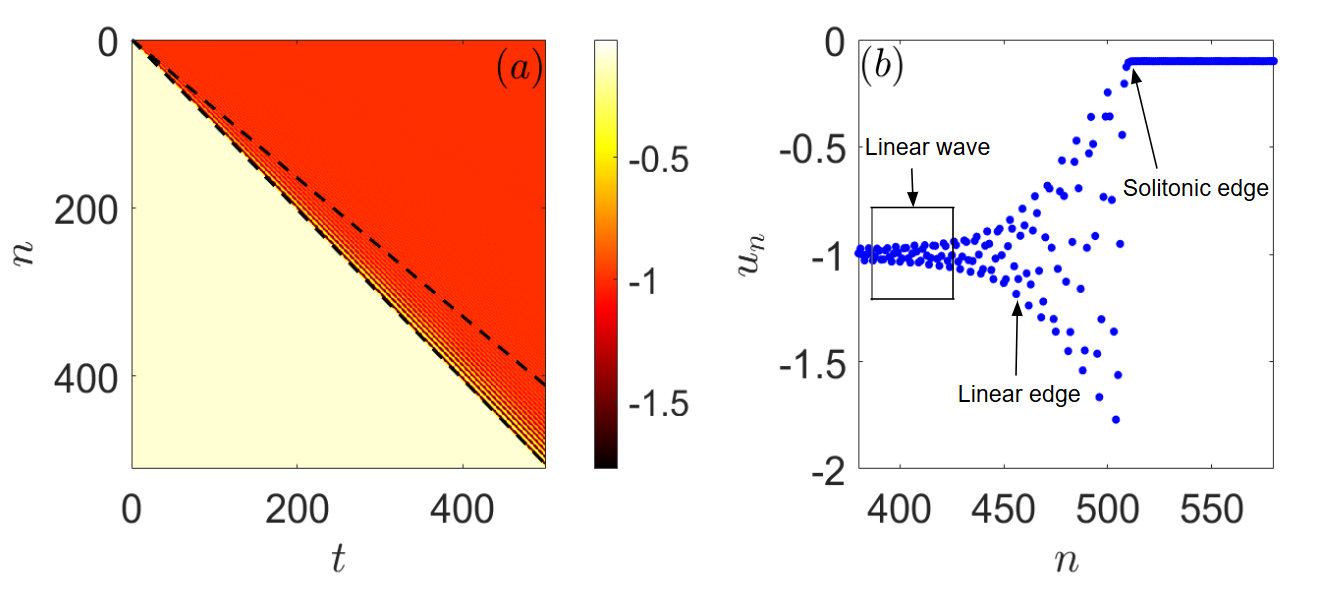}
    \caption{The discrete dispersive shock wave of the lattice \eqref{eq: granular crystals}. The left panel $(a)$ shows the space-time dynamics of the DDSW, where the two dashed black lines represent the theoretical predictions on the linear and solitonic edges of the DDSW based on the DSW-fitting results (See Section \ref{sec: DSW fitting}), while the right panel $(b)$ depicts the spatial profile of the lattice DDSW at $t = 500$. Notice that $\alpha = 1.1$.}
    \label{fig:DDSW}
\end{figure}

\begin{figure}[t!]
    \centering
    \includegraphics[width=0.9\linewidth]{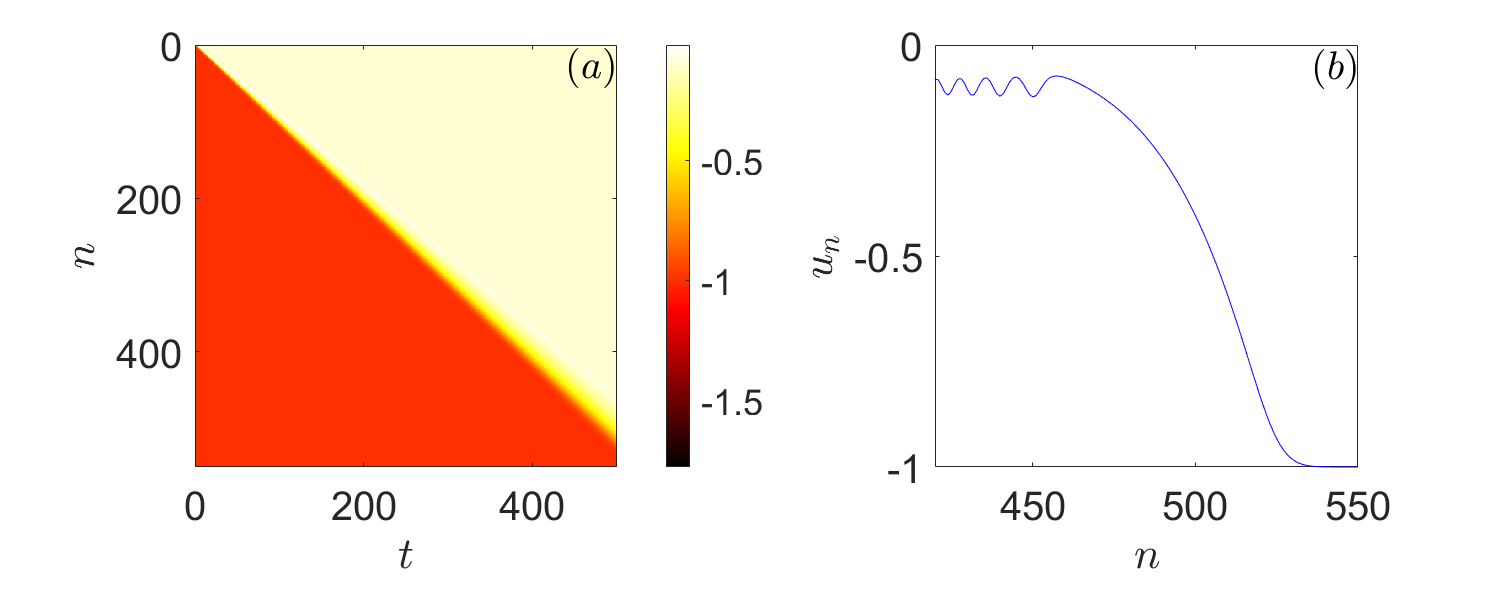}
    \caption{The discrete rarefaction wave. The panel $(a)$ and $(b)$ refer to the density plot of the magnitude of the field $u_n$ and the spatial profile of the discrete RW of the lattice \eqref{eq: granular crystals} at $t = 500$, respectively. Also, note that $\alpha = 1.1$.}
    \label{fig:DRW plot}
\end{figure}

\section{Model description and theoretical setup}\label{sec: model intro}

In this work, we study the so-called Fermi-Pasta-Ulam (FPU) lattice whose equation of motion can be described as the following second-order differential-difference equations (DDEs),
\begin{equation}\label{eq: original lattice}
    \frac{d^2x_n} {dt^2} = V'\left(x_{n+1} - x_n\right) - V'\left(x_n - x_{n-1}\right), 
\end{equation}
where $n \in \mathbb{Z}$, $x_n(t) \in \mathbb{R}$ refers to the displacement of the $n$th particle at time $t$, and $V$ is a potential function governing the pairwise interaction between the particles.

For technical convenience, it is more preferable to write the DDEs in Eq.~\eqref{eq: original lattice} in terms of the relative displacement (aka. strains) $u_n = x_n - x_{n-1}$ so that Eq.~\eqref{eq: original lattice} becomes,
\begin{equation}\label{eq: lattice at the strain level}
    \frac {d^2u_n}{dt^2} = V'(u_{n+1}) - 2V'(u_n) + V'(u_{n-1}).
\end{equation}
In this paper, we are interested in the following homogeneous fully nonlinear potential function,
\begin{equation}\label{eq: Hertzian pot}
    V(x) = \frac{1}{1+\alpha}\left|x\right|^{1+\alpha},
\end{equation}
where $\alpha > 1$.

With such potential function of $V$ in Eq.~\eqref{eq: Hertzian pot}, the equation of motion at the level of the relative displacement can be further written as follows,
\begin{equation}\label{eq: granular crystals}
    \frac {d^2u_n}{dt^2} = \Delta\left(u_n|u_n|^{\alpha-1}\right),
\end{equation}
where $u_n(t) \leq 0$, $\alpha > 1$ denotes a parameter, and $\Delta\left[f\right] \equiv f_{n+1} - 2f_{n} + f_{n-1}$ is the discrete Laplace operator. We note that in this paper, without loss of generality, we are only interested in the scenario where $u_n(t)$ is negative for all $n$.

The lattice \eqref{eq: granular crystals} shall be main focus of this present work, and we claim it is a dispersive model. To see this, we look for a plane wave solution in the form of $u_n(t) = \overline{u} + a\text{e}^{i(kn-\omega t)}$ and we substitute it into the lattice \eqref{eq: granular crystals} to obtain the following two-branch linear dispersion relation, upon eliminating the small parameter $0<a\ll1$,
\begin{equation}\label{eq: linear dr of the lattice}
    \omega(\overline{u},k) = \pm 2\sqrt{\alpha}|\overline{u}|^{\frac{\alpha-1}{2}}\sin\left(\frac{k}{2}\right).
\end{equation}
Obviously, the linear dispersion relation \eqref{eq: linear dr of the lattice} does not always have zero derivative with respect to the wavenumber $k$, so the lattice in Eq.~\eqref{eq: granular crystals} is a discrete dispersive model. Since the main goal of this work is to study the DSWs of the lattice \eqref{eq: granular crystals}, we equip the DDEs in \eqref{eq: granular crystals} with the following so-called Riemann initial data,
\begin{equation}\label{eq: Riemann ICs}
    u_n(0) = \begin{cases}
        u_-, \quad n \leq 0,\\
        u_+, \quad n > 0,
    \end{cases}
    \quad
    v_n(0) = \begin{cases}
        v_-, \quad n \leq 0,\\
        v_+, \quad n > 0.
    \end{cases}
\end{equation}
where $v_n(t) := du_n/dt$, and $u_{\pm}, v_{\pm}$ are four real constants referring to the homogeneous background values. Figures \ref{fig:DDSW} and \ref{fig:DRW plot} depict the dynamics of the lattice \eqref{eq: granular crystals} subject to the ICs \eqref{eq: Riemann ICs}, where an initial upward jump leads to the DSW, while a downward jump yields a RW.

Next, we review two quasi-continuum models proposed in Ref.~\cite{10.1098/rspa.2013.0462}. To this end, as $\alpha \to 1^+$, we introduce the following change of variables,
\begin{equation}\label{eq: change of variables}
   \begin{aligned}
    &u_n(t) \sim v(\xi,\tau); \\
    &\xi = 2\sqrt{3}\epsilon\left(n - t\right), \quad \tau = \sqrt{3}\epsilon^3 t,
    \end{aligned}
\end{equation}
where $0 < \epsilon \ll 1$ is a formal smallness parameter which depends on $\alpha$ and is to be determined.

Now setting $\epsilon = \sqrt{\alpha - 1} \ll 1$, and substituting the ansatz \eqref{eq: change of variables} into the Eq.~\eqref{eq: granular crystals} yields the following log-KdV equation \cite{10.1098/rspa.2013.0462}, by collecting terms at the order of $\mathcal{O}(\epsilon^4)$,
\begin{equation}\label{eq: log-kdv approximation}
    v_\tau + \left(v\ln|v|\right)_\xi + v_{\xi\xi\xi} = 0,
\end{equation}
We then compute the linear dispersion relation of Eq.~\eqref{eq: log-kdv approximation} by assuming the following plane-wave ansatz: 
\begin{equation}\label{eq: plane-wave ansatz}
v(\xi,\tau) = \overline{v} + a\exp\left[i(k\xi-\omega\tau)\right], 
\end{equation}
where $0<a\ll1$ denotes a smallness parameter. We note that the plane-wave in Eq.~\eqref{eq: plane-wave ansatz} has the background $\overline{v} \in \mathbb{R}$.

A direct substitution of Eq.~\eqref{eq: plane-wave ansatz} into the log-KdV equation eq.~\eqref{eq: log-kdv approximation} yields the following linear dispersion relation, upon eliminating $a$,
\begin{equation}\label{eq: log-kdv ldr}
    \omega(\overline{v},k) = \left(1+\ln\left|\overline{v}\right|\right) k- k^3.
\end{equation}

Moreover, a generalized KdV equation with H\"older-continuous nonlinearity can also be derived as an approximation for Eq.~\eqref{eq: granular crystals}:
\begin{equation}\label{eq: generalized KdV}
    v_\tau + \frac{\alpha}{\alpha-1}\left(v - v|v|^{1/\alpha-1}\right)_\xi + v_{\xi\xi\xi} = 0.
\end{equation}
For Eq.~\eqref{eq: generalized KdV}, we similarly compute its linear dispersion relation by plugging the plane-wave ansatz \eqref{eq: plane-wave ansatz} into Eq.~\eqref{eq: generalized KdV}, and eliminating the smallness parameter $a$ to obtain that,
\begin{equation}\label{eq: gkdv ldr}
    \omega(\overline{v},k) = \frac{\alpha}{\alpha-1}\left(1-\frac{1}{\alpha}\left|\overline{v}\right|^{1/\alpha-1}\right)k - k^3.
\end{equation}

In work \cite{10.1098/rspa.2013.0462}, the two quasi-continuum models are proposed to approximate the discrete solitary wave of the granular crystal lattice \eqref{eq: granular crystals}. However, since our purpose is to apply them to approximate the discrete DSW numerically observed in the lattice \eqref{eq: granular crystals}, it is hence still necessary to conduct such throughout reviews on the derivations of these relevant models. In addition, we observe that the two quasi-continuum models in Eqs.~\eqref{eq: log-kdv approximation} and \eqref{eq: generalized KdV}, in principle, are valid whenever $\alpha$ is close to $1$. Regarding this observation, we mainly concentrate on the scenario of $\alpha = 1.1$ throughout this work.

\section{Traveling solitary-wave solutions}\label{sec: solitary waves}

In this section, we review the solitary-wave solutions of the two models of Eq.~\eqref{eq: log-kdv approximation} and \eqref{eq: generalized KdV}. To this end, we assume the solitary-wave solution in the form,
\begin{equation}\label{eq: TW ansatz}
    v(\xi,\tau) = V(z), \quad z = \xi - c\tau,
\end{equation}
where $c$ is a real constant referring to the speed of propagation of the traveling solitary wave. 

\begin{figure}[t!]
    \centering
    \includegraphics[width=0.8\linewidth]{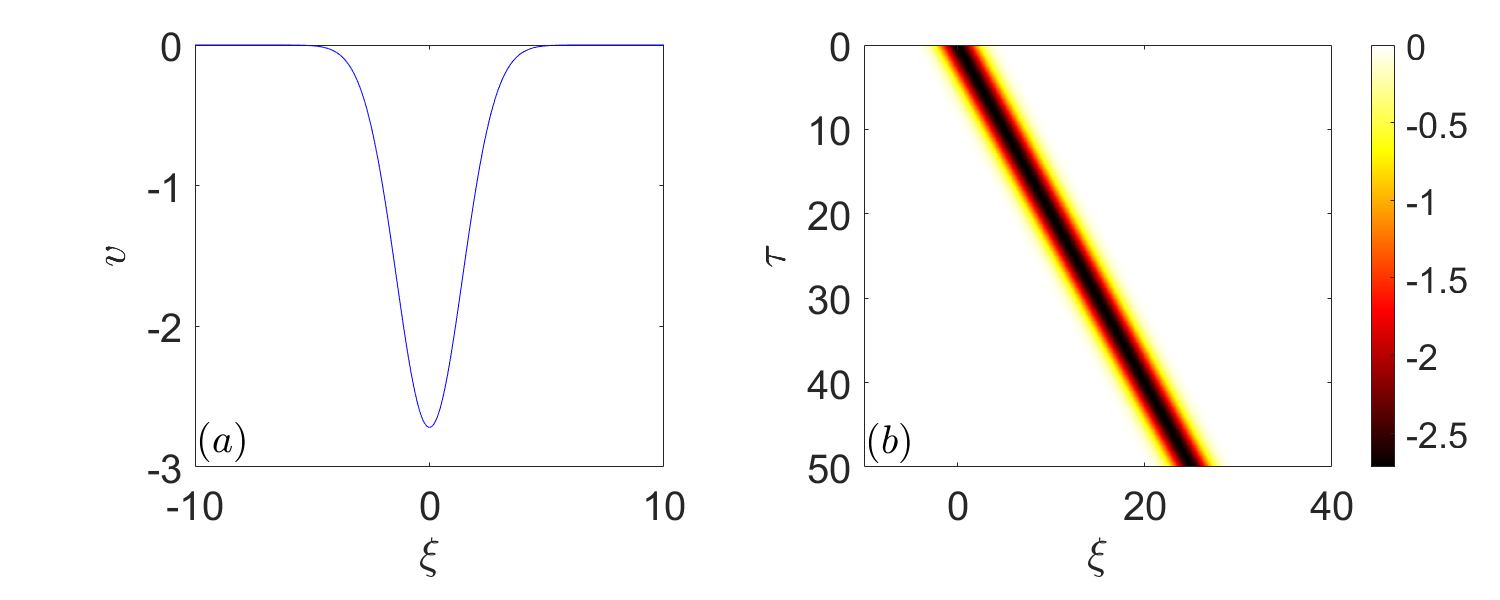}
    \caption{The Gaussian "dark" solitary wave in Eq.~\eqref{eq: solitart-wave for log-kdv}. The left panel $(a)$ depicts the spatial profile of the dark solitary wave at $\tau = 0$, while the right panel $(b)$ displays the space-time evolution dynamics of such solitary wave with a speed of propagation $c = 0.5$.}
    \label{fig:Gaussian soliton}
\end{figure}

\paragraph{Log-KdV solitary wave:} Substituting the anstaz in Eq.~\eqref{eq: TW ansatz} into Eq.~\eqref{eq: log-kdv approximation} yields,
\begin{equation}\label{eq: TW ODE}
    -cV_z + \left(V\ln|V|\right)_z + V_{zzz} = 0.
\end{equation}
Assuming that $\lim_{z\to\pm\infty}V(z) = 0$, a direct integrating of Eq.~\eqref{eq: TW ODE} yields the following traveling solitary-wave solution \cite{10.1098/rspa.2013.0462},
\begin{equation}\label{eq: solitart-wave for log-kdv}
    V(z) = -\text{e}^{1/2+c}\text{e}^{-z^2/4}.
\end{equation}
Figure \ref{fig:Gaussian soliton} shows the spatial profile of this Gaussian-profile solitary wave \eqref{eq: solitart-wave for log-kdv} and also its associated space-time evolution dynamics.

\paragraph{Generalized-KdV solitary wave:} On the other hand, for the generalized-KdV equation \eqref{eq: generalized KdV}, substitution of the ansatz \eqref{eq: TW ansatz} into Eq.~\eqref{eq: generalized KdV} yields,
\begin{equation}\label{eq: TW ODE for gkdv}
    -cV_z + \frac{\alpha}{\alpha-1}\left(V-V|V|^{1/\alpha-1}\right)_z + V_{zzz} = 0.
\end{equation}
Again we suppose $\lim_{z\to\pm\infty}V(z) = 0$, and then we integrate Eq.~\eqref{eq: TW ODE for gkdv} to obtain the following solitary-wave solution \cite{10.1098/rspa.2013.0462},
\begin{equation}\label{eq: solitary-wave for gkdv}
    V(z) = -\left(1-\mu\right)^{\alpha/(1-\alpha)}F\left[\sqrt{1-\mu}z\right]; \quad \mu = c\left(1-1/\alpha\right),
\end{equation}
where
\begin{equation}
    F(\zeta) = \begin{cases}
        \widetilde{A}\cos^{2\alpha/(\alpha-1)}\left(\widetilde{B}\zeta\right); \quad |\zeta| \leq \frac{\pi}{2\widetilde{B}},\\
        0, \quad |\zeta| \geq \frac{\pi}{2\widetilde{B}},
    \end{cases}
\end{equation}
and
\begin{equation}
    \widetilde{A} = \left(\frac{1+\alpha}{2\alpha}\right)^{\alpha/(1-\alpha)},\quad \widetilde{B} = \frac{\sqrt{\alpha-1}}{2\sqrt{\alpha}}.
\end{equation}

We finally end this section of reviewing the solitary-wave solutions with a note. It is very important to realize that these solitary-wave solutions (Eqs.~\eqref{eq: solitart-wave for log-kdv} and \eqref{eq: solitary-wave for gkdv}) play an essential role in the DSW as it is expected that the profile of the solitonic edge of the DSW ought to agree with those of their associated solitary waves.

\section{Periodic solutions}\label{sec: periodic solutions}

In this section, we discuss briefly the existence of the periodic solutions to the two quasi-continuum models. To the best of knowledge, it is very unfortunate that these periodic solutions are not amenable to analytical treatment. However, it shall be useful to visualize the potential curves of the associated co-traveling frame ODE to understand their existence.

On the one hand, for the log-KdV equation, upon integrating the Eq.~\eqref{eq: TW ODE}, we obtain that,
\begin{equation}\label{eq: first-order ode for log-kdv}
    \left(V_z\right)^2 = \left(c-\ln|V|+\frac{1}{2}\right)V^2 + 2AV + B,
\end{equation}
where $A,B$ are two constants of integration.

Analogously, for the generalized-KdV equation \eqref{eq: generalized KdV}, we integrate Eq.~\eqref{eq: TW ODE for gkdv} to arrive at,
\begin{equation}\label{eq: first-order ode for gkdv}
    \left(V_z\right)^2 = \left(c-\frac{\alpha}{\alpha-1}\right)V^2 + \frac{2\alpha^2}{(\alpha-1)(\alpha+1)}|V|^{1/\alpha+1} + 2EV + F,
\end{equation}
where $E,F$ are two constants of integration.

We shall visualize the potential curves associated with Eqs.~\eqref{eq: first-order ode for log-kdv}-\eqref{eq: first-order ode for gkdv} to analyze the type of distinct traveling-wave solutions of the log-KdV \eqref{eq: log-kdv approximation} and generalized-KdV \eqref{eq: generalized KdV} equations. In particular, the potential curves are:
   \begin{align}
    &P_{\text{log-KdV}}(V) = -\left(c-\ln|V|+\frac{1}{2}
    \right)V^2-2AV-B,\label{eq: Potential curves of log-kdv}\\
    &P_{\text{g-KdV}}(V) = -\left(c-\frac{\alpha}{\alpha-1}\right)V^2 - \frac{2\alpha^2}{(\alpha-1)(\alpha+1)}\left|V\right|^{1/\alpha+1}-2EV-F. \label{eq: potential curve of g-kdv}
    \end{align}
\begin{figure}[t!]
    \centering
    \includegraphics[width=0.7\linewidth]{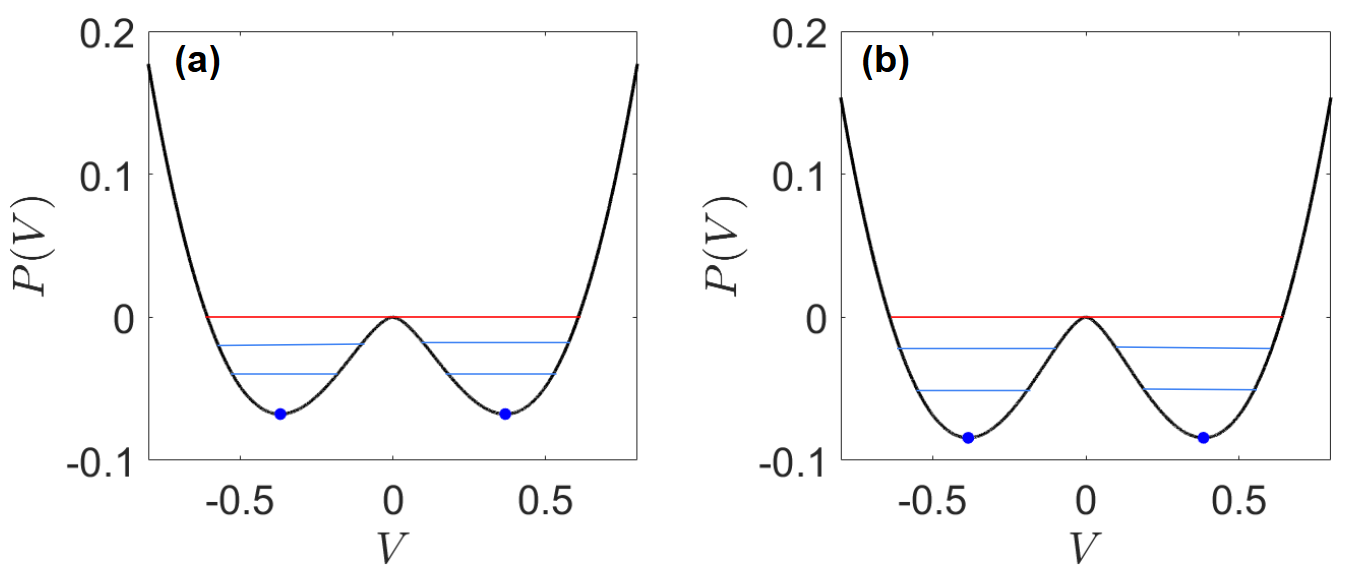}
    \caption{The potential curves. Panel $(a)$ depicts the potential curve \eqref{eq: Potential curves of log-kdv} of the log-KdV equation \eqref{eq: log-kdv approximation}, while $(b)$ shows that \eqref{eq: potential curve of g-kdv} of the generalized-KdV equation \eqref{eq: generalized KdV}.}
    \label{fig:Potential curves}
\end{figure}
Figure \ref{fig:Potential curves} displays the two potential curves in Eqs.~\eqref{eq: Potential curves of log-kdv}-\eqref{eq: generalized KdV}. Specifically, we observe that the horizontal red lines represent the homoclinic orbits which shall associate with those solitary-wave solutions that we reviewed in Section \ref{sec: solitary waves}. In addition, the blue lines in panels $(a)$ and $(b)$ depict the periodic orbits and they are corresponding to the periodic traveling-wave solutions to the two quasi-continuum models of Eqs.~\eqref{eq: log-kdv approximation} and \eqref{eq: generalized KdV}.

\section{Conservation laws and Lagrangian densities}\label{sec: conservation laws}

In this section, we list some important conservation laws of the two quasi-continuum models. The purpose of mentioning these necessary conservation laws is that they shall be the prerequisite of the derivation of Whitham modulation equations. In particular, we will utilize the so-called averaging the conservation laws approach (see section \ref{sec: Whitham theory} for details) for such derivation. Meanwhile, we mention, in addition, the corresponding Lagrangian densities of the two quasi-continuum models since they can also be very useful in the derivation of the modulation system despite through another method entitled to averaging the Lagrangian. 

\paragraph{Log-KdV conservation laws:} For the log-KdV equation, it admits the following conservation of \textit{mass} and \textit{momentum},
\begin{equation}\label{eq: log-kdv conservation laws}
    \begin{aligned}
        &v_\tau + \left(v\ln\left|v\right| + v_{\xi\xi}\right)_\xi = 0,\\
        &\frac{1}{2}\left(v^2\right)_\tau + \left(vv_{\xi\xi} + \frac{v^2}{4}+\frac{1}{2}v^2\ln\left|v\right|-\frac{1}{2}\left(v_\xi\right)^2\right)_\xi = 0,
    \end{aligned}
\end{equation}

\paragraph{generalized-KdV conservation laws:} For the generalized-KdV equation,
\begin{equation}\label{eq: gkdv conservation laws}
    \begin{aligned}
        &v_\tau + \left[\frac{\alpha}{\alpha-1}\left(v - v\left|v\right|^{1/\alpha-1}\right) + v_{\xi\xi}\right]_\xi = 0,\\
        &\frac{1}{2}\left(v^2\right)_\tau + \left[vv_{\xi\xi} + \frac{\alpha}{\alpha-1}\left(\frac{1}{2}v^2-\frac{1}{\alpha+1}\left|v\right|^{1/\alpha+1}\right)-\frac{1}{2}\left(v_\xi\right)^2\right]_\xi = 0.
    \end{aligned}
\end{equation}
In addition, it is relevant to notice that the two quasi-continuum models of Eqs.~\eqref{eq: log-kdv approximation} and \eqref{eq: generalized KdV} conserve the energy as well:
\begin{equation}\label{eq: energy conservations}
    E(v) = \frac{1}{2}\int_{\mathbb{R}}\left(v_\xi\right)^2 \text{d}\xi - \int_{\mathbb{R}}W(v)\text{d}\xi,
\end{equation}
where $W(v) = \frac{v^2}{2}\left(\ln|v|-\frac{1}{2}\right)$ for Eq.~\eqref{eq: log-kdv approximation}, and $W(v) = \frac{\alpha}{\alpha-1}\left(\frac{1}{2}v^2-\frac{\alpha}{\alpha+1}|v|^{1/\alpha+1}\right)$ for Eq.~\eqref{eq: generalized KdV}. However, since we will not need the energy conservation law for the derivation of the Whitham modulation equations (See Section \ref{sec: Whitham theory}), we omit the associated fluxes for brevity.

Furthermore, we notice that the two quasi-continuum models Eqs.~\eqref{eq: log-kdv approximation} and \eqref{eq: generalized KdV} both admit Lagrangian structures with the following closed form Lagrangian densities,
\begin{equation}\label{eq: Lagrangian densities}
\begin{aligned}
    &\mathbb{L} = \frac{1}{2}\psi_\xi\psi_\tau + \frac{1}{2}\ln\left|\psi_\xi\right|\left(\psi_\xi\right)^2 - \frac{1}{4}\left(\psi_\xi\right)^2 - \frac{1}{2}\left(\psi_{\xi\xi}\right)^2,\\
    &\mathbb{L} = \frac{1}{2}\psi_\xi\psi_\tau + \frac{\alpha}{2\left(\alpha-1\right)}\left(\psi_\xi\right)^2-\frac{\alpha^2}{(\alpha-1)(\alpha+1)}\left|\psi_\xi\right|^{1/\alpha+1} - \frac{1}{2}\left(\psi_{\xi\xi}\right)^2,
\end{aligned}
\end{equation}
respectively.

\section{Whitham modulation equations}\label{sec: Whitham theory}

We now derive the Whitham modulation equations which shall govern the slowly-varying spatial and temporal dynamics of the parameters of the periodic solutions of the quasi-continuum models. Firstly, it is important to observe that both quasi-continuum models of Eqs.~\eqref{eq: log-kdv approximation} and \eqref{eq: generalized KdV} are PDEs of order $3$, so this suggests that there are three parameters in the periodic solutions of both models. Hence, we need three equations to form a closed modulation system which governs the slow spatial and temporal evolution dynamics of those parameters. To compute the modulation equations, we first assume a slowly modulated wave in the following form,
\begin{equation}\label{eq: slowly varied periodic wave}
    v(\xi,\tau) = \varphi(\theta;X,T) + \delta v_1\left(\theta;X,T\right) + \mathcal{O}(\delta^2), \quad 0 < \delta \ll 1,
\end{equation}
where $X = \delta \xi$ and $T = \delta \tau$ are two slow variables, and $\theta = S(X,T)/\epsilon$ is an associated fast phase that satisfies, via the chain rule,
\begin{equation}\label{eq: fast-phase definition}
    \theta_\xi = S_X = k(X,T), \quad \theta_\tau = S_T = -\omega(X,T).
\end{equation}
We also note that the functions $\varphi(\theta;X,T)$ and $v_1(\theta;X,T)$ in Eq.~\eqref{eq: slowly varied periodic wave} are assumed to be periodic solutions with a fixed spatial period with respect to the fast phase of $\theta$. Based on the section \ref{sec: periodic solutions}, we know the analytical form of the periodic solutions are not obtainable, but they exist, of course, due to the phase-plane analysis performed (See Figure \ref{fig:Potential curves}). Hence, it is reasonable to expect that in principle, one has the freedom to fix the period of the periodic solution of $\varphi(\theta)$ to any arbitrary value via a process of reparametrization (See \cite{Hoefer:2009} for an example on the reparametrization of the periodic solution of the standard KdV equation). For this reason, we assume the period of $\varphi(\theta)$ to be $2\pi$ in $\theta$.

What follows immediately from the definition of the fast phase $\theta$ in Eq.~\eqref{eq: fast-phase definition} is that the compatibility condition of $S_{XT} = S_{TX}$ leads to the following equation,
\begin{equation}\label{eq: conservation of waves}
    k_T + \omega_X = 0,
\end{equation}
which is the so-called \textit{conservation of waves}, and it is the first modulation equation within the closed modulation system. To find the remaining modulation equations, we shall apply the method of averaging the conservation laws \cite{10.1063/1.1947120,EL201611}. To this end, we first define the following average operation: For a given function $F$,
\begin{equation}\label{eq: Average operation}
    \overline{F(\phi)} = \frac{1}{2\pi}\int_0^{2\pi}F(\phi)\text{d}\theta.
\end{equation}

\paragraph{Log-KdV modulation equations:} For the log-KdV equation \eqref{eq: log-kdv approximation}, we apply the average operation defined in Eq.~\eqref{eq: Average operation} to the two conservation laws in Eq.~\eqref{eq: log-kdv conservation laws} and collect terms on the distinct order of $\delta$ to obtain that,
\begin{equation}\label{eq: modulations for log-kdv}
    \begin{aligned}
    &\left(\overline{\phi}\right)_T + \left(\overline{\phi\ln\left|\phi\right|}\right)_X = \mathcal{O}(\delta),\\
    &\left(\overline{\phi^2}\right)_T + \left(\frac{1}{2}\overline{\phi^2}+\overline{\phi^2\ln\left|\phi\right|}-3k^2\overline{\left(\phi_\theta\right)^2}\right)_X = \mathcal{O}(\delta).
    \end{aligned}
\end{equation}
We then drop all the terms at the order of $\mathcal{O}(\delta)$ and higher to arrive at,
\begin{equation}\label{eq: final closed modulation system for log-kdv}
    \begin{aligned}
    &k_T + \omega_X = 0,\\
    &\left(\overline{\phi}\right)_T + \left(\overline{\phi\ln\left|\phi\right|}\right)_X = 0,\\
    &\left(\overline{\phi^2}\right)_T + \left(\frac{1}{2}\overline{\phi^2} + \overline{\phi^2\ln\left|\phi\right|} - 3k^2\overline{\left(\phi_\theta\right)^2}\right)_X = 0.
    \end{aligned}
\end{equation}

\paragraph{Generalized-KdV modulation equations:} Similarly, for the generalized-KdV equation \eqref{eq: generalized KdV}, we applied the average operation to its conservation laws in Eq.~\eqref{eq: gkdv conservation laws} to obtain at, at the order of $\mathcal{O}(\delta)$,
\begin{equation}\label{eq: modulations for generalized-kdv}
    \begin{aligned}
    &\left(\overline{\phi}\right)_T + \frac{\alpha}{\alpha-1}\left(\overline{\phi} - \overline{\phi\left|\phi\right|^{1/\alpha-1}}\right)_X = \mathcal{O}(\delta),\\
    &\left(\overline{\phi^2}\right)_T + \left[\frac{2\alpha}{\alpha-1}\left(\overline{\phi^2}-\frac{1}{\alpha+1}\overline{\left|\phi\right|^{1/\alpha-1}}\right)-3k^2\overline{\left(\phi_\theta\right)^2}\right]_X = \mathcal{O}(\delta)
    \end{aligned}
\end{equation}
Ignoring terms at the order of $\mathcal{O}(\delta)$ then yields,
\begin{equation}\label{eq: final closed modulation system for gkdv}
    \begin{aligned}
        &k_T + \omega_X = 0,\\
        &\left(\overline{\phi}\right)_T + \frac{\alpha}{\alpha-1}\left(\overline{\phi}-\overline{\phi\left|\phi\right|^{1/\alpha-1}}\right)_X = 0,\\
        &\left(\overline{\phi^2}\right)_T + \left[\frac{2\alpha}{\alpha-1}\left(\overline{\phi^2}-\frac{1}{\alpha+1}\overline{\left|\phi\right|^{1/\alpha-1}}\right) - 3k^2\overline{\left(\phi_\theta\right)^2}\right]_X = 0.
    \end{aligned}
\end{equation}
We note that the system \eqref{eq: final closed modulation system for log-kdv} and \eqref{eq: final closed modulation system for gkdv} represent the final complete closed system of modulation equations for the log-KdV \eqref{eq: log-kdv approximation} and generalized-KdV \eqref{eq: generalized KdV} model, respectively. Meanwhile, we should end this section with a few comments. Firstly, it is worthwhile to mention that averaging the conservation laws is not the only methodology for the derivation of the Whitham modulation equations, one can also apply the method of multiple scale expansion \cite{EL201611} or averaging the Lagrangian \cite{Yang_Biondini_Chong_Kevrekidis_2025}. But, these three methods are expected to be equivalent in the sense that the final modulation system derived through either method should coincide. For an illustrative purpose, we also apply the method of averaging the Lagrangian to derive the full modulation system, and we leave the details to appendix \ref{sec: Appendix} for interested readers.

\section{Reduction of the modulation equations}\label{sec: Whitham reductions}

In this section, we discuss how the Whitham modulation system derived in section \ref{sec: Whitham theory} can be reduced to the so-called simple-wave ordinary differential equation which encodes the important information on the edge features of the DSW at both its linear and solitonic edge. We note, however, that such reduction can only be performed at either the harmonic or the solitonic limit of the modulation equations. Since the derivation of the simple-wave ODE is similar for both modulation system of the two quasi-continuum models, we only show the details of the derivation for the log-KdV modulation system.

At the harmonic limit, it is expected that the frequency $\omega$ simply asymptotes to the corresponding linear dispersion relation, denoted as $\omega_0$, of the model. Moreover, the latter two modulation equations in Eq.~\eqref{eq: final closed modulation system for log-kdv} shall reduce into one equation, so that the modulation system now becomes,
\begin{equation}\label{eq: modulation system at the harmonic limit}
   \begin{aligned}
    &k_T + (\omega_0)_X = 0,\\
    &\left(\overline{\phi}\right)_T + \left(\overline{\phi}\ln\left|\overline{\phi}\right|\right)_X = 0.
    \end{aligned}
\end{equation}
We note that, to obtain the system \eqref{eq: modulation system at the harmonic limit}, we have also two facts at the harmonic limit: $(a)$ The term $\overline{\left(\phi_\theta\right)^2}$ asymptotes to $0$ and $(b)$:
\begin{equation}
    \overline{F\left(\phi\right)} = F\left(\overline{\phi}\right).
\end{equation}
These two facts can attribute to the small amplitude oscillation of the wave, which occurs at the linear edge of the DSW (See \cite{10.1063/1.1947120} for more elaborations on the two facts). Then, we cast the system \eqref{eq: modulation system at the harmonic limit} into the following matrix form,
\begin{equation}\label{eq: matrix form modulation equation}
   \begin{bmatrix}
       k \\
       \phi
   \end{bmatrix}_T + 
   \begin{bmatrix}
       \partial_k\omega_0 & \partial_{\overline{\phi}}\omega_0\\
       0 & 1 + \ln\left|\overline{\phi}\right|
   \end{bmatrix}
   \begin{bmatrix}
       k \\
       \overline{\phi}
   \end{bmatrix}_X = \vec{0},
\end{equation}
and notice that coefficient matrix possesses the left eigenpair: $\left(\lambda,\vec{v}\right)$, where
\begin{equation}
    \lambda = \partial_k\omega_0, \quad \vec{v} = \left[1+\ln\left|\overline{\phi}\right|-\partial_k\omega_0, -\partial_{\overline{\phi}}\omega_0\right].
\end{equation}
We then multiply this left eigenvector on both sides of Eq.~\eqref{eq: matrix form modulation equation} to obtain the following characteristic form,
\begin{equation}
    \left(1+\ln\left|\overline{\phi}\right|-\partial_k\omega_0\right)\frac{dk}{dT} - \partial_{\overline{\phi}}\omega_0\frac{d\overline{\phi}}{dT} = 0,
\end{equation}
which can be further simplified into,
\begin{equation}\label{eq: final simple-wave ODE}
    \frac{dk}{d\overline{\phi}} = \frac{\partial_{\overline{\phi}}\omega_0}{1+\ln\left|\overline{\phi}\right|-\partial_k\omega_0}.
\end{equation}
Eq.~\eqref{eq: final simple-wave ODE} is the so-called simple-wave ODE mentioned before, and we moreover notice that it needs to be equipped with the condition that $k\left(v_+\right) = 0$ since the linear-edge wavenumber at the solitonic edge of the DSW has to be $0$. Thus, the simple-wave ODE in Eq.~\eqref{eq: final simple-wave ODE} with the constraint that $k(v_+) = 0$ forms an initial-value problem.

On the other hand, at the soliton limit of the modulation system, it will be much more computationally friendly to introduce the following conjugate dispersion relation \cite{10.1063/1.1947120},
\begin{equation}\label{eq: conjugate dispersion relation}
    \widetilde{\omega}_s(\overline{\phi},\widetilde{k}) = -i\omega_0\left(\overline{\phi},i\widetilde{k}\right),
\end{equation}
where $\widetilde{k}$ denotes the conjugate wavenumber. Then, a similar argument leads to the following simple-wave ODE,
\begin{equation}\label{eq: simple-wave ode at soliton limit}
    \frac{d\widetilde{k}}{d\overline{\phi}} = \frac{\partial_{\overline{\phi}}\widetilde{\omega}_s}{1+\ln\left|\overline{\phi}\right| - \partial_{\widetilde{k}}\widetilde{\omega}_s}, \quad \widetilde{k}(v_-) = 0.
\end{equation}

As a final note before we end this section, we shall solve these two IVPs in Eq.~\eqref{eq: final simple-wave ODE} and \eqref{eq: simple-wave ode at soliton limit} in section \ref{sec: DSW fitting} in order to gain insights and theoretical predictions on various DSW edge features such as the linear and solitonic egde speeds. The reductions of the Whitham modulation system at the harmonic and soliton limits is entitled to the \textit{DSW fitting} method.

\section{Rarefaction waves}\label{sec: RWs}
In this section, we analytically compute the self-similar solutions associated with the two quasi-continuum models in Eqs.~\eqref{eq: log-kdv approximation} and \eqref{eq: generalized KdV} which shall serve as approximations to the rarefaction waves observed numerically in the granular lattice \eqref{eq: granular crystals}. To this end, we assume the following self-similar ansatz,
\begin{equation}\label{eq: self-similar ansatz}
    v(\xi,\tau) = S(\kappa), \quad \kappa = \frac{\xi}{\tau}.
\end{equation}

\paragraph{Log-KdV RW:}For the log-KdV equation \eqref{eq: log-kdv approximation}, we notice that its dispersionless limit reads,
\begin{equation}\label{eq: dispersionless limit of log-kdv}
    v_\tau + \left(v\ln|v|\right)_\xi = 0,
\end{equation}
which is subject to the downward step initial condition:
\begin{equation}
    v(\xi,0) = 
    \begin{cases}
        v_-,\quad \xi \leq 0\\
        v_+,\quad \xi > 0,
    \end{cases}
\end{equation}
where $v_- > v_+$.

A direct substitution of the self-similar ansatz of Eq.~\eqref{eq: self-similar ansatz} into Eq.~\eqref{eq: dispersionless limit of log-kdv} yields,
\begin{equation}\label{eq: self-similar ODE}
    -\kappa S_\kappa + \left(S\ln|S|\right)_\kappa = 0.
\end{equation}
Then, solving Eq.~\eqref{eq: self-similar ODE} for $S$ yields the following self-similar solution,
\begin{equation}\label{eq: self-similar solution}
    v(\xi,\tau) = \begin{cases}
        v_-; \quad \xi \leq \left(1+\ln|v_-|\right)\tau,\\
        -\exp\left(\xi/\tau-1\right); \quad \left(1+\ln|v_-|\right)\tau < \xi \leq \left(1+\ln|v_+|\right)\tau,\\
        v_+; \quad \xi > \left(1+\ln|v_+|\right)\tau
    \end{cases}
\end{equation}
To compare analytical self-similar solution in Eq.~\eqref{eq: self-similar solution} with the discrete RW of the granular lattice Eq.~\eqref{eq: granular crystals}, we transform the spatial and temporal coordinates $(\xi,\tau)$ back into $(n,t)$ so that,
\begin{equation}
   \begin{aligned}
    u_n(t) = v(n,t) = 
    \begin{cases}
        v_-, \quad n \leq b_-(t),\\
        -\exp\left(\frac{2(n-t)}{\epsilon^2t} - 1\right), \quad b_-(t) < n \leq b_+(t),\\
        v_+, \quad n > b_+(t),
    \end{cases}
    \end{aligned}
\end{equation}
where
\begin{equation}
    b_\pm(t) = t + \frac{1}{2}\epsilon^2t\left(1+\ln|v_\pm|\right).
\end{equation}

\paragraph{Generalized-KdV RW:} Similarly, for the generalized-KdV equation, we notice that its corresponding dispersionless system is given as follows,
\begin{equation}\label{eq: dispersionless generalized KdV}
    v_\tau + \frac{\alpha}{\alpha-1}\left(v - v|v|^{1/\alpha-1}\right)_\xi = 0.
\end{equation}
Then, substitution of the ansatz Eq.~\eqref{eq: self-similar ansatz} into Eq.~\eqref{eq: dispersionless generalized KdV} yields,
\begin{equation}\label{eq: self-similar ODE for gkdv}
    -\kappa S_\kappa + \frac{\alpha}{\alpha-1}\left(S-S|S|^{1/\alpha-1}\right)_\kappa = 0.
\end{equation}
Solving for $S$ in Eq.~\eqref{eq: self-similar ODE for gkdv} yields,
\begin{equation}\label{eq: self-similar solution of gkdv}
    v(\xi,\tau) = \begin{cases}
        v_-, \quad \xi \leq \frac{\alpha-|v_-|^{(1-\alpha)/\alpha}}{\alpha-1}\tau,\\
        -\left[\alpha-\left(\alpha-1\right)\frac{\xi}{\tau}\right]^{\alpha/(1-\alpha)}, \quad \frac{\alpha-|v_{-}|^{(1-\alpha)/\alpha}}{\alpha-1}\tau < \xi \leq \frac{\alpha-|v_+|^{(1-\alpha)/\alpha}}{\alpha-1}\tau,\\
        v_+, \quad \xi > \frac{\alpha-|v_+|^{(1-\alpha)/\alpha}}{\alpha-1}\tau.
    \end{cases}
\end{equation}
We then transform the self-similar solution in Eq.~\eqref{eq: self-similar solution of gkdv} to the spatial and temporal coordinates of $(n,t)$ to obtain that,
\begin{equation}
    u_n(t) = \begin{cases}
        v_-; \quad n \leq c_-(t),\\
        -\left[\alpha-\frac{2(\alpha-1)(n-t)}{\epsilon^2t}\right]^{\alpha/(1-\alpha)}; \quad c_-(t) < n \leq c_+(t),\\
        v_+; \quad n > c_+(t), 
    \end{cases}
\end{equation}
where 
\begin{equation}
    c_{\pm}(t) = t + \frac{1}{2}\epsilon^2t\left(\frac{\alpha-|v_\pm|^{(1-\alpha)/\alpha}}{\alpha-1}\right).
\end{equation}

\begin{figure}[t!]
    \centering
    \includegraphics[width=0.8\linewidth]{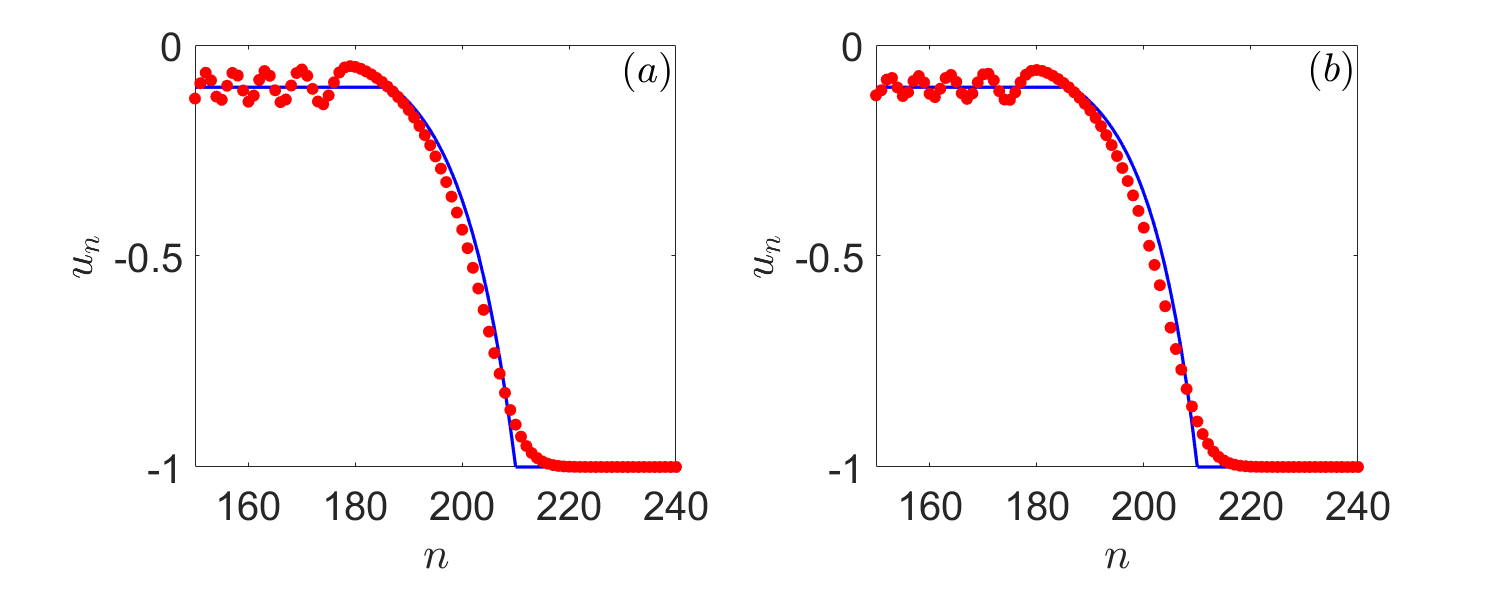}
    \caption{The comparisons of the rarefaction waves: Panels $(a)$ and $(b)$ depicts the comparison of the RWs between the log-KdV, generalized-KdV self-similar solutions in Eqs.~\eqref{eq: self-similar solution}, \eqref{eq: self-similar solution of gkdv} and the associated lattice RWs of the discrete model \eqref{eq: granular crystals} at $t = 200$, where we utilize the parameter value $\alpha = 1.1$.} 
    \label{fig:RW comparisons}
\end{figure}

\section{DSW fitting}\label{sec: DSW fitting}

In this section, we apply the methodology of DSW fitting in order to gain some insights on the edge features of the DSW of the quasi-continuum models. In particular, we shall make analytical and quantitative predictions on the DSW characteristics including the edge speeds and wavenumber.

The DSW-fitting method \cite{10.1063/1.1947120} essentially boils down to a system of \textit{simple-wave} ordinary differential equations (ODEs) which encode the information of the wavenumber at the two edges of the DSW. On the one hand, we recall from section \ref{sec: Whitham reductions} that near the linear edge of the DSW, the wavenumber $k$ shall satisfy the following initial-value problem (IVP) which is reiterated here for convenience,
\begin{equation}\label{eq: linear-edge ode}
    \frac{dk}{d\overline{v}} = \frac{\partial_{\overline{v}}\omega_0}{A(\overline{v}) - \partial_k\omega_0}, \quad k(v_+) = 0.
\end{equation}
where $\omega_0$ denotes the linear dispersion relation of the model, and we will specify the definition of $A(\overline{v})$ later.

On the other hand, we also recall from section \ref{sec: Whitham reductions} that near the solitonic edge of the DSW, the so-called conjugate wavenumber, denoted by $\widetilde{k}$, should fulfill the following simple-wave ODE,
\begin{equation}\label{eq: solitonic-edge ode}
    \frac{d\widetilde{k}}{d\overline{v}} = \frac{\partial_{\overline{v}}\widetilde{\omega}_s}{A(\overline{v}) - \partial_{\widetilde{k}}\widetilde{\omega}_s}, \quad \widetilde{k}(v_-) = 0,
\end{equation}
where $\widetilde{\omega}_s(\overline{v},\widetilde{k}) = -i\omega_0\left(\overline{v},i\widetilde{k}\right)$ refers to the conjugate dispersion relation.

With the explicit linear dispersion relations in Eqs.~\eqref{eq: log-kdv ldr} and \eqref{eq: gkdv ldr} of the two quasi-continuum models, we solve the two IVPs in Eqs.~\eqref{eq: linear-edge ode}-\eqref{eq: solitonic-edge ode} to first gain analytical predictions on the linear-edge wavenumber which is denoted as $k^-$.

\paragraph{Log-KdV DSW fitting:} For the log-KdV equation \eqref{eq: log-kdv approximation}, we note first that $A(\overline{v}) = 1 + \ln|\overline{v}|$, and we recall that its linear dispersion relation is given in Eq.~\eqref{eq: log-kdv ldr}. Then, solving the IVP in Eq.~\eqref{eq: linear-edge ode} yields,
\begin{equation}
    k(\overline{v}) = \sqrt{\frac{2}{3}\ln\left(\frac{|\overline{v}|}{|v_+|}\right)}.
\end{equation}
Hence, the linear-edge wavenumber $k_-$ should be given as:
\begin{equation}\label{eq: linear-edge wavenumber}
    k_- = k\left(v_-\right) = \sqrt{\frac{2}{3}\ln\left(\frac{|v_-|}{|v_+|}\right)}.
\end{equation}
Furthermore, we compute the group velocity at $k_-$ to obtain the linear-edge velocity of the DSW, denoted by $s_-$,
\begin{equation}\label{eq: linear-edge speed of log-kdv dsw}
   \begin{aligned}
    s_- = \partial_k\omega_0\left(v_-,k_-\right)
    = 1 + \ln|v_-| - 3k_-^2.
    \end{aligned}
\end{equation}

On the other hand, near the solitonic-edge of the DSW, we solve the IVP specified in equation \eqref{eq: solitonic-edge ode} with the conjugate dispersion relation:
\begin{equation}\label{eq: log-kdv conjugate dr}
    \widetilde{\omega}_s(\overline{v},\widetilde{k}) = \left(1+\ln|\overline{v}|\right)\widetilde{k} + \widetilde{k}^3.
\end{equation}
A direction integration of the IVP in Eq.~\eqref{eq: solitonic-edge ode} yields,
\begin{equation}\label{eq: conjugate wavenumber}
    \widetilde{k}(\overline{v}) = \sqrt{\frac{2}{3}\ln\left(\frac{|v_-|}{|\overline{v}|}\right)}.
\end{equation}
Then, the solitonic-edge speed, denoted by $s_+$, is obtained by computing the phase velocity,
\begin{equation}\label{eq: solitonic-edge speed of log-kdv DSW}
    \begin{aligned}
        s_+ = \frac{\widetilde{\omega}_s}{\widetilde{k}}\left(v_+,\widetilde{k}_+\right)
        = 1 + \ln|v_+| + \widetilde{k}_+^2.
    \end{aligned}
\end{equation}
where $\widetilde{k}_+ = \widetilde{k}(v_+)$.

\paragraph{Generalized-KdV DSW fitting:} For the DSW of the generalized-KdV equation \eqref{eq: generalized KdV}, we analogously follow the procedures which we performed for the log-KdV DSW fitting. Namely, Noticing that $A(\overline{v}) = \frac{\alpha}{\alpha-1}\left(1-\frac{1}{\alpha}|\overline{v}|^{1/\alpha-1}\right)$, and the conjugate dispersion relation:
\begin{equation}\label{eq: gkdv conjugate dr}
    \widetilde{\omega}_s(\overline{v},\widetilde{k}) = \frac{\alpha}{\alpha-1}\left(1-\frac{1}{\alpha}|\overline{v}|^{1/\alpha-1}\right)\widetilde{k} + \widetilde{k}^3.
\end{equation}
Solving the two IVPs in Eqs.~\eqref{eq: linear-edge ode} and \eqref{eq: solitonic-edge ode} yields,
\begin{equation}\label{eq: gkdv wavenumber}
    \begin{aligned}
        k &= \sqrt{\frac{2}{3(\alpha-1)}\left(-|\overline{v}|^{1/\alpha-1} + |v_+|^{1/\alpha-1}\right)},\\
        \widetilde{k} &= \sqrt{\frac{2}{3(\alpha-1)}\left(\left|\overline{v}\right|^{1/\alpha-1}-\left|v_-\right|^{1/\alpha-1}\right)}.
    \end{aligned}
\end{equation}
Then, the linear and solitonic-edge speeds of the generalized-KdV DSW is given as follows,
\begin{equation}\label{eq: gkdv dsw edge speeds}
    \begin{aligned}
        s_- = \partial_k\omega_0\left(v_-,k_-\right) = \frac{\alpha}{\alpha-1}\left(1-\frac{1}{\alpha}\left|v_-\right|^{1/\alpha-1}\right) -3k_-^2,\\
        s_+ = \frac{\widetilde{\omega}_s}{\widetilde{k}}\left(v_+,\widetilde{k}_+\right) = \frac{\alpha}{\alpha-1}\left(1-\frac{1}{\alpha}\left|v_+\right|^{1/\alpha-1}\right) + \widetilde{k}_+^2.
    \end{aligned}
\end{equation}
where now $\omega_0$ is given in Eq.~\eqref{eq: gkdv ldr}, and $k_- = k(v_-), \widetilde{k}_+ = \widetilde{k}(v_+)$.

We notice that these DSW-fitting theoretical predictions on the edge speeds (Eqs.~\eqref{eq: linear-edge speed of log-kdv dsw}, \eqref{eq: solitonic-edge speed of log-kdv DSW} and \eqref{eq: gkdv dsw edge speeds}) of the DSW of the two quasi-continuum models will be compared to the numerically measured DSW edge speeds in the later section to examine the performance of the DSW-fitting analysis on the DSW edge characteristics. To this end, we have to transform these DSW-fitting theoretical predictions on the DSW edge speeds in Eqs.~\eqref{eq: linear-edge speed of log-kdv dsw}, \eqref{eq: solitonic-edge speed of log-kdv DSW} and \eqref{eq: gkdv dsw edge speeds} into the speeds associated in the $(n,t)$ spatial and temporal coordinate, and this can be achieved by the following observation,
\begin{equation}\label{eq: speeds transform}
    \frac{dn}{dt} = \frac{\alpha-1}{2}\frac{d\xi}{d\tau} + 1.
\end{equation}

\section{Numerical validation}\label{sec: Numerical validation}

We are now ready to compare the theoretical predictions on different edge features (e.g. edge speeds) of the DSWs of the quasi-continuum models with those which will be measured numerically based on the time evolution dynamics of the quasi-continuum models and the discrete lattice. To begin with, we discuss, in detail, some indispensable numerical setup which include the integration schemes and the approach to compute the edge speeds of the DSWs numerically, etc...

\subsection{Integration schemes and initial conditions}
\begin{figure}[t!]
    \centering
    \includegraphics[width=0.4\linewidth]{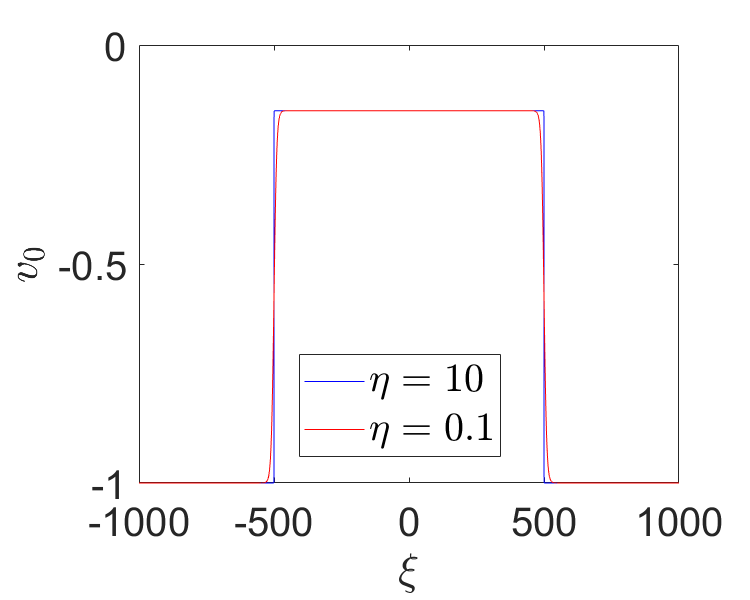}
    \caption{The box-type initial conditions in Eq.~\eqref{eq: ICs for quasi-continuum models} for two distinct smoothing parameters of $\eta$, where the two location parameters are $a = -500$ and $b = 500$.}
    \label{fig:ICs}
\end{figure}

Firstly, for the discrete lattice in Eq.~\eqref{eq: granular crystals}, we apply a fourth-order Runge-Kutta (RK4) method for time integration with Neumann boundary conditions to solve the differential-difference equations \eqref{eq: granular crystals}. Next, for the two quasi-continuum models in Eqs.~\eqref{eq: log-kdv approximation} and \eqref{eq: generalized KdV}, we utilize the exponential time differencing RK$4$ (ETDRK$4$) time integration scheme \cite{doi:10.1137/S1064827502410633} with a pseudo-spectral method of spatial discretization to perform time stepping of the two PDEs. Next, for the initial conditions of the quasi-continuum models, we consider the following "box-type" initial data,
\begin{equation}\label{eq: ICs for quasi-continuum models}
    v_0(\xi) = v(\xi,0) = v_+ - \frac{1}{2}\left(v_+-v_-\right)\left[\tanh\left(\eta\left(\xi-a\right)\right) - \tanh\left(\eta\left(\xi - b\right)\right)\right],
\end{equation}
where $\eta > 0$ is a constant characterizing the smoothness of the initial jumps, and $a,b\in\mathbb{R}$ are two parameters which decide the left and right location of the initial jump, respectively. Specifically, one should expect that as $\eta$ decreases, the initial jumps shall become smoother. The main purpose of producing a smooth IC is to avoid the potential modulational instability which can possibly occur when one uses an IC with a sharp jump as it associates with high-frequency components in Fourier space. Although one reason to apply the IC in Eq.~\eqref{eq: ICs for quasi-continuum models} is to have a smoother version of the standard Riemann initial data which has a sharp jump, another key reason is that the pseudo-spectral discretization method assumes the periodic boundary conditions. Figure \ref{fig:ICs} Shows two ICs with different values of $\eta$, and we can clearly see that when $\eta = 10$, the IC has a jump which is much shaper than that corresponding to $\eta = 0.1$.

One the one hand, to compare the dynamics of the lattice \eqref{eq: granular crystals} with that of the log-KdV equation \eqref{eq: log-kdv approximation}, we need to use the following initial condition for the velocity $s_n \equiv \dot u_n$,
\begin{equation}\label{eq: Initial velocity}
    s_n(0) = -2\sqrt{3}\epsilon v_\xi\left(\xi,0\right) - \sqrt{3}\epsilon^3\left[\left(v(\xi,0)\ln|v(\xi,0)|\right)_\xi + v_{\xi\xi\xi}(\xi,0)\right],
\end{equation}
where $\xi = 2\sqrt{3}\epsilon n$.

On the other hand, to compare the dynamics of the lattice \eqref{eq: granular crystals} with that of the generalized KdV equation \eqref{eq: generalized KdV}, we need 
\begin{equation}
    s_n(0) = -2\sqrt{3}\epsilon v_{\xi}(\xi,0) - \sqrt{3}\epsilon^3\left(\frac{\alpha}{\alpha-1}\left(v(\xi,0) - v(\xi,0)|v(\xi,0)|^{1/\alpha - 1}\right)_\xi + v(\xi,0)_{\xi\xi\xi}\right).
\end{equation}

\begin{figure}[b!]
    \centering
    \includegraphics[width=0.8\linewidth]{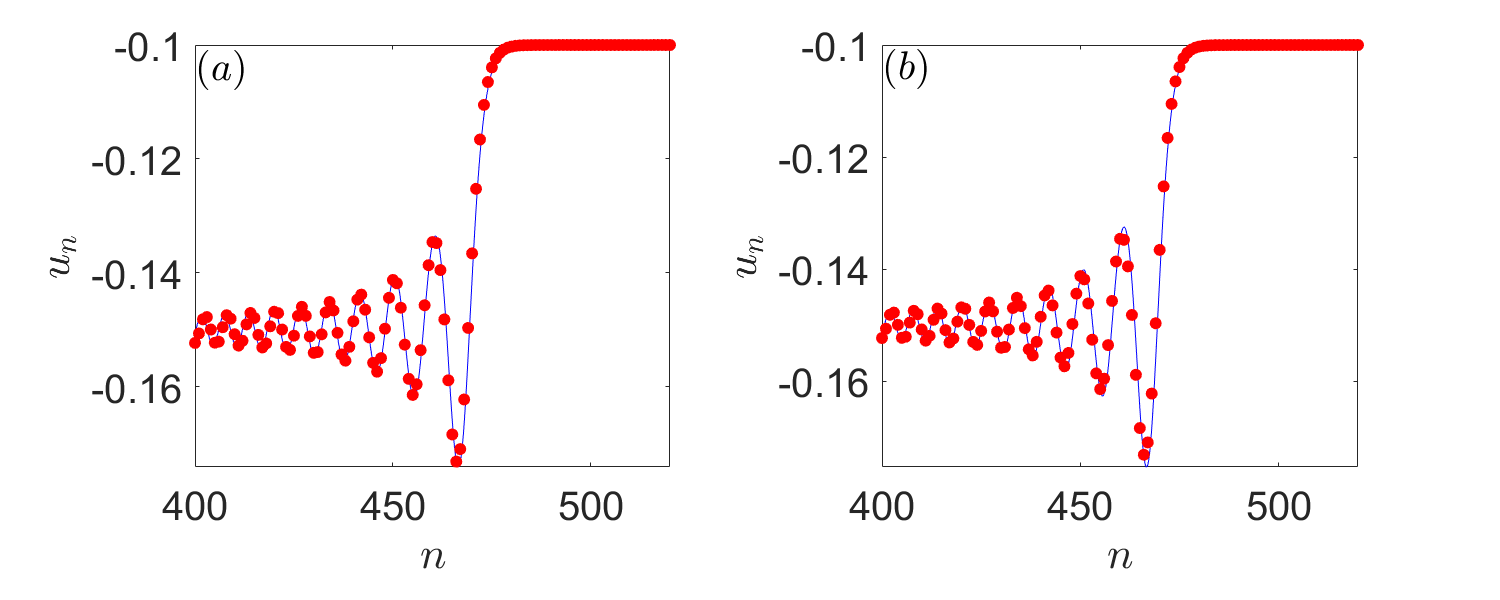}
    \caption{The comparison of the DSWs: Panel $(a)$ and $(b)$ depict the comparison of the DDSW of the lattice \eqref{eq: granular crystals} (red dots) with the DSW (blue curves) of the log-KdV Eq.~\eqref{eq: log-kdv approximation} and of the generalized-KdV \eqref{eq: generalized KdV} at $t = 500$, respectively. Notice that the parameter values are $\alpha = 1.1$, $u_- = -0.15$, and $u_+ = -0.1$.}
    \label{fig:DSW comparisons}
\end{figure}

\subsection{Numerical estimation of DSW edge speeds}

We then discuss briefly the method utilized to numerically measure the solitonic and linear-edge speeds of the DSWs of both quasi-continuum models of Eqs.~\eqref{eq: log-kdv approximation} and \eqref{eq: generalized KdV} and the lattice \eqref{eq: granular crystals}. On the one hand, for the solitonic-edge speed of the DSWs which is denoted as $s_+$, we first compute the global minimizer of the field $v(\xi,t)$ at a given time $t$. Namely, we find $\xi_+ = \text{argmin}_\xi v(\xi,t)$, and we simply treat such $\xi_+$ as the solitonic location of the DSWs. To numerically measure the solitonic-edge speed of $s_+$, we fix two time snapshots, say $t_1$ and $t_2$. Then, we compute the two solitonic locations, denoted as $\xi_+^1$ and $\xi_+^2$, of the DSW at these two $t$. The numerical solitonic-edge speed is then given as follows,
\begin{equation}\label{eq: numerical s_+}
    s_+ = \left|\frac{\xi_+^2 - \xi_+^1}{t_2 - t_1}\right|.
\end{equation}
Finally, since the formula in Eq.~\eqref{eq: numerical s_+} yields a $s_+$ in the coordinate of $(\xi,\tau)$, one shall transform it into an associated $s_+$ in the $(n,t)$ coordinate by applying Eq.~\eqref{eq: speeds transform}. On the other hand, for the numerical linear-edge speed which is denoted as $s_-$, we still pick two time snapshots of $t_1$ and $t_2$. Then, we try to pinpoint the linear-edge location of the DSWs as follows: we first define the following two quantities,
\begin{equation}\label{eq: window cores}
    I^{U} = u_- + \frac{\left|u_--u_+\right|}{N}, \quad I^{L} = u_- - \frac{\left|u_--u_+\right|}{N},
\end{equation}
where $N$ is a positive integer. Next, we compute local maximum and minimum of the DSW which belong to the two windows of $\left(I^{U}-\widetilde{\epsilon}, I^{U} + \widetilde{\epsilon}\right)$ and $\left(I^{L}-\widetilde{\epsilon}, I^{L} + \widetilde{\epsilon}\right)$. For these local extremes, we apply a least-square based method to fit two straight lines, and these two lines shall intersect at a point, say $\xi_-$. We then simply regard such $\xi_-$ as the linear-edge location of the DSWs. We shall obtain $\xi_-^1$ and $\xi_-^2$ at two distinct time snapshots of $t_1$ and $t_2$, and we apply the formula in Eq.~\eqref{eq: numerical s_+} with $\xi_+$ replaced by $\xi_-$ to numerically evaluate the linear-edge speed of $s_-$ of the DSWs.

\subsection{Numerical comparisons}\label{subsec: numerical comparison}

\begin{figure}
    \centering
    \includegraphics[width=0.8\linewidth]{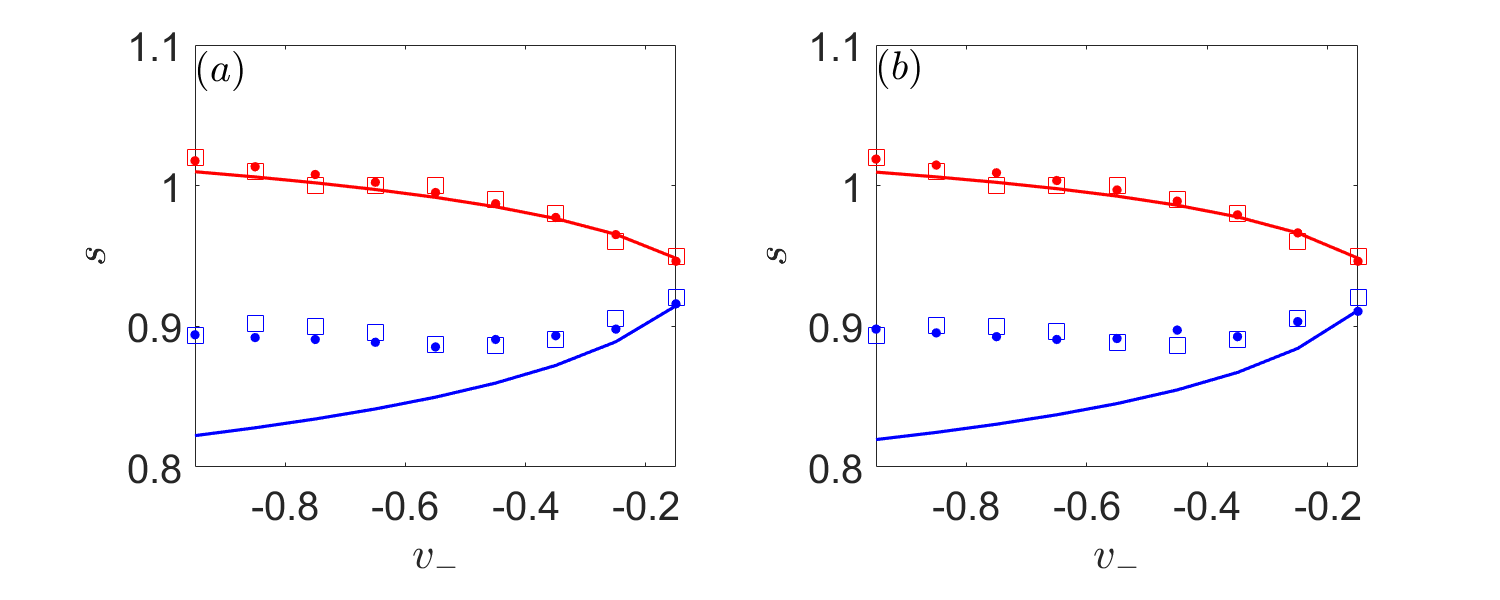}
    \caption{The DSW edge-speed comparisons. The left panel $(a)$ shows the DSW edge speeds of the log-KdV \eqref{eq: log-kdv approximation} DSW and that of the associated lattice \eqref{eq: granular crystals}, while the right panel $(b)$ depicts the comparison between the generalized-KdV and lattice. Notice that the red curves, dots, and squares represent the DSW-fitting theoretical predictions, numerical quasi-continuum and lattice DSWs solitonic-edge speeds, and the associated blue ones refer to DSWs linear-edge speeds. Finally, notice that the background value of $v_+$ is fixed to be $-0.1$, and $\alpha = 1.1$, as usual.}
    \label{fig:edge-speeds comparison}
\end{figure}

We are now ready to perform comparisons in order to examine how well the two quasi-continuum models approximate the DDSW of the lattice \eqref{eq: granular crystals}. Firstly, figure \ref{fig:DSW comparisons} displays the comparison of the DSW of the lattice with that of the associated quasi-continuum model at a specific time snapshot $t = 500$. In particular, we can clearly see that the DDSWs which are represented as red dots agree quite well with the DSW of the quasi-continuum models, which is depicted as blue curves. However, one may expect that such good agreement is due to the fact that the jump of Riemann initial data, denoted as $\Delta = \left|u_- - u_+\right|$ is small, which is $\Delta = 0.05$ in the current scenario. Hence, the discrepancy may become more prominent as the value of jump $\Delta$ becomes larger. Indeed, figure \ref{fig:edge-speeds comparison} displays the DSW edge-speed comparisons. The panel $(a)$ shows the comparison between the log-KdV \eqref{eq: log-kdv approximation} and the associated lattice \eqref{eq: granular crystals}, while the panel $(b)$ analogously shows that for the generalized-KdV \eqref{eq: generalized KdV} with the corresponding lattice \eqref{eq: granular crystals}. Based on these two comparisons, we can clearly conclude that the DSW-fitting theoretical predictions on the DSW edge speeds agree better as $\Delta$ decreases. But, even though the jump is relatively large (e.g. $\Delta = 0.8$), it is still evident that the solitonic-edge speeds agree well simultaneously with a much more prominent deviation on the linear-edge speeds. However, one may observe the wiggling feature of the data points of the linear-edge speeds. This is expected due to the uncertainty of measuring the linear-edge location. In particular, for example, using a different $N$ in Eq.~\eqref{eq: window cores} can yield a slight variation of the linear-edge location. 

\subsection{Influence of the parameter $\alpha$}

\begin{figure}[b!]
    \centering
    \includegraphics[width=0.8\linewidth]{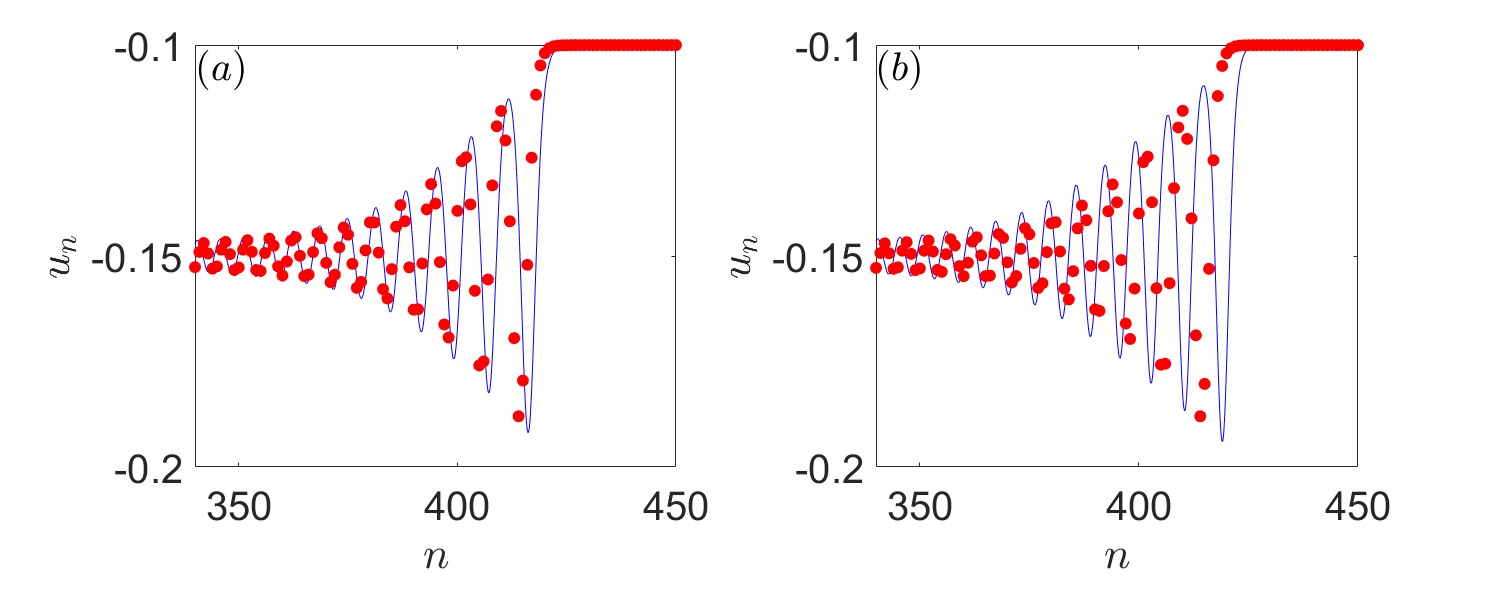}
    \hfill
    \includegraphics[width=0.8\linewidth]{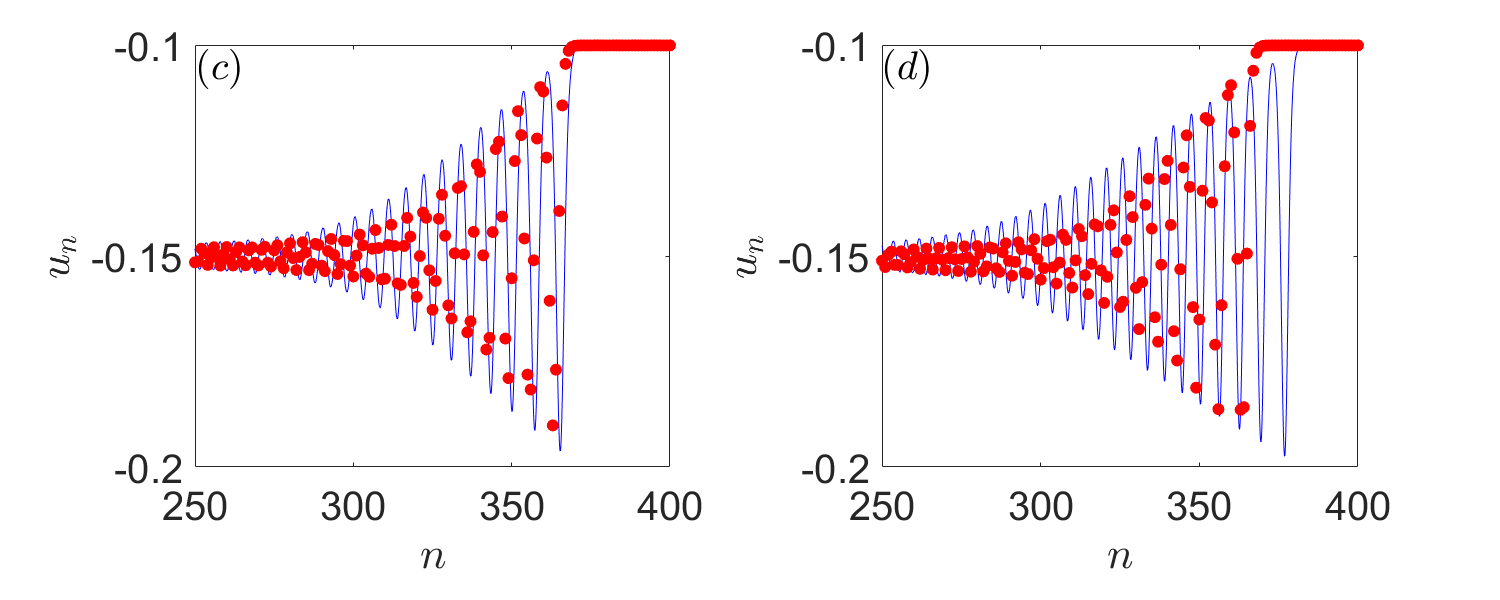}
    \caption{The comparison of the DSWs associated with other values of $\alpha$. The two panels on the first row associate with $\alpha = 1.3$, while the second row corresponds to $\alpha = 1.5$. Similarly, the left panels $(a)$ and $(c)$ refer to the comparison of the discrete DSW of the lattice \eqref{eq: granular crystals} with that of the log-KdV equation \eqref{eq: log-kdv approximation}, while the right panels $(b)$ and $(d)$ represent the comparison of the discrete DSW with the DSW of the generalized-KdV equation \eqref{eq: generalized KdV}. Note that the values of the two homogeneous backgrounds are $u_- = -0.15$ and $u_+ = -0.1$, and relevant comparisons are made at the specific time snapshot $t = 500$.}
    \label{fig: The comparison of the DSW: other alpha's}
\end{figure}

We then discuss briefly how the parameter of $\alpha$ can affect the approximations of the two quasi-continuum models to the lattice of \eqref{eq: granular crystals}. As we have already seen in subsection \ref{subsec: numerical comparison}, the comparison between the quasi-continuum DSW with that of the lattice is good as long as we have a sufficiently small initial jump of $\Delta$ with the fixed $\alpha = 1.1$. However, with a same jump of $\Delta$, if we increase the value of $\alpha$ from the previously fixed $1.1$, we shall expect a worse comparison as the quasi-continuum models of Eqs.~\eqref{eq: log-kdv approximation} and \eqref{eq: generalized KdV} are expected to be valid only if $\alpha \to 1^+$, or equivalently, $\epsilon = \sqrt{\alpha-1}$ is small. To further illustrate this expectation, we compare the DSW of both quasi-continuum models and that of the lattice \eqref{eq: granular crystals} for other values of $\alpha$ including $1.3$ and $1.5$. Meanwhile, we note that the scenario that $\alpha = 1.5$ should gain the most interest as the lattice in Eq.~\eqref{eq: granular crystals} is actually entitled to the granular crystal lattice. Figure \ref{fig: The comparison of the DSW: other alpha's} displays the comparisons of the DSWs associated with other two values of $\alpha$: $1.3$ and $1.5$. It is clear that the comparisons become worse especially for the comparison between the discrete DSW of the lattice \eqref{eq: granular crystals} and that of the generalized-KdV \eqref{eq: generalized KdV}. For example, the solitonic edge of the discrete DSW of lattice deviates that of the quasi-continuum models, indicating the prominent difference of the solitonic-edge speeds between the discrete DSW and the quasi-continuum DSWs. However, once again, we recall that the two quasi-continuum models are supposed to be valid for $\alpha$ which is sufficiently close to $1$. With the values of $\alpha$ getting farer away from $1$, we cannot expect a equally good comparison as the situation when $\alpha = 1.1$, and therefore we conclude that, via the numerical validation, the two quasi-continuum models fail to provide good approximations to the discrete DSW structure of the lattice \eqref{eq: granular crystals}.

\section{Conclusions and future challenges}\label{sec: conclusions}

In this paper, we reviewed two previously established quasi-continuum models which serve as approximations of the granular crystal lattice. We studied their analytical traveling solitary-wave solutions as they are important to demonstrate the solitonic egde profile of the DSW. The periodic solutions of the two quasi-continuum models, although not amenable to analytical treatment, were shown to exist by our detailed phase-plane analysis. Based on these periodic solutions, we derive the complete and closed Whitham modulation system for both quasi-continuum models, which governs the slowly varying spatial and temporal dynamics of all the relevant parameters of the periodic traveling waves. We further performed an indispensable reduction on the Whitham modulation system and obtained a simple-wave ODE that actually encodes some crucial information on some edge features of the DSWs such as the edge wavenumber and speeds. We then compared these theoretical predictions with their associated numerical counterparts and thereafter verify that both quasi-continuum models indeed provide reasonable approximations of the DSWs to that of the lattice. Finally, we also studied and approximated the rarefaction waves of the lattice via computing the analytical self-similar solutions of the dispersionless correspondences of the quasi-continuum models, and compared it with the numerically observed RWs from the lattice, and such comparison also exhibits good agreement.

However, there are still many open questions and we here only list two of them. The first one is regarding the Whitham modulation systems for the quasi-continuum models. In particular, although we have successfully derived the closed modulation system for both models, the question is how explicitly each modulation system is closed? To the best of our knowledge, this problem unfortunately cannot be resolved under any analytical approaches mainly because the periodic solutions do not admit an analytical form. Hence, it is impossible to know what precise parameters of the periodic solutions to be expected, although it may quite likely include the amplitude, wavenumber, and/or speed of propagation of the periodic traveling wave. This issue can perhaps be bypassed via applying a data-driven discovery of the Whitham modulation equations. Namely, one can attempt to utilize some machine-learning based (e.g. SINDy \cite{doi:10.1073/pnas.1517384113}, neural ODE \cite{chen2019neuralordinarydifferentialequations}, and/or Koopman method \cite{doi:10.1137/21M1401243,KORDA2020599}) method to perform numerical identification of the Whitham modulation equations as a substitute of Eqs.~\eqref{eq: final closed modulation system for log-kdv} and \eqref{eq: final closed modulation system for gkdv} which are all expressed with the variable of wave mean $\overline{\phi}$. Instead, the identified equations shall not necessarily be expressed in terms of the wave mean $\overline{\phi}$, but the modulation equations may can be expressed in terms of other essential parameters of the periodic solutions such as the wave amplitude and traveling speed. Secondly, we notice that the two quasi-continuum models in Eqs.~\eqref{eq: log-kdv approximation} and \eqref{eq: generalized KdV} are all first-order (in time) uni-directional models. The work \cite{10.1098/rspa.2013.0462} proposed a second-order (in time) bi-directional quasi-continuum model for the lattice \eqref{eq: granular crystals}, so it shall be very interesting to perform a similar study with such bi-directional model and investigate how it can be used to study the RWs and DSWs in the lattice \eqref{eq: granular crystals}. Methodology related to bi-directional model is present in \cite{EL201611} and can be applied for the bi-directional model, as well. All these open directions are currently under investigations and will be reported in future publications. 

\section*{Appendix}\label{sec: Appendix}

In this Appendix, we derive, in detail, the Whitham modulation equations for the log-KdV equation by the method of averaging the Lagrangian for an illustrative purpose. To this end, we notice first that for periodic solutions of the log-KdV equation \eqref{eq: log-kdv approximation}, the behavior of the newly defined field $\psi$ shall generally be,
\begin{equation}\label{eq: a larangian ansatz}
    \psi = \beta \xi - \gamma \tau + \varphi(\theta), \quad \theta = k\xi - \omega \tau,
\end{equation}
where $\beta,\gamma$ are two arbitrary real constants, and the function $\varphi(\theta)$ is assumed to be periodic with a fixed spatial period $2\pi$ and with a zero average, namely $\overline{\psi} = 0$. 

For such phase of $\theta$, the traveling-wave ansatz of $V(\theta)$ now satisfies,
\begin{equation}
    k^2V_\theta^2 = \left(c_p+\frac{1}{2}\right)V^2 - V^2\ln\left|V\right| + 2AV + B \equiv G(V),
\end{equation}
where $c_p = \omega/k$ is defined as the phase speed, and $A,B$ are two constants of integration.

Importantly, it is worthwhile to note that the parameter pair of $(\beta,\gamma)$ is a counterpart to $(k,\omega)$ and we call $\vartheta = \beta\xi - \gamma\tau$ the \textit{pseudo-phase} \cite{Yang_Biondini_Chong_Kevrekidis_2025}. We note that $V(\theta)$ relates to $\psi\left(\xi,\tau\right)$ and its associated derivatives as follows,
\begin{equation}\label{eq: key relations}
    \begin{aligned}
        &\psi_\xi = \beta + k\varphi_\theta = V,\\
        &\psi_\tau = -\gamma - c_p\left(V-\beta\right),\\
        &\psi_{\xi\xi} = kV_\theta.
    \end{aligned}
\end{equation}
Then, a direct substitution of Eq.~\eqref{eq: key relations} into the Lagrangian density in \eqref{eq: Lagrangian densities} yields,
\begin{equation}\label{eq: substitution of L density}
   \begin{aligned}
    \mathbb{L} &= -k^2V_\theta^2 + \left(A-\frac{1}{2}\gamma+\frac{1}{2}c_p\beta\right)V + \frac{1}{2}B .
    \end{aligned}
\end{equation}
We compute the average Lagrangian,
\begin{equation}\label{eq: Averaged Lagrangian}
    \mathcal{L} = \frac{1}{2\pi}\int_0^{2\pi}\mathbb{L}\text{d}\theta = kW\left(A,B,c_p\right) + \left(A-\frac{1}{2}\gamma+\frac{1}{2}c_p\beta\right)\beta + \frac{1}{2}B,
\end{equation}
where the so-called "action integral" $W(A,B,c_p)$ is defined as follows,
\begin{equation}
    W(A,B,c_p) = -\frac{1}{2\pi}\oint\left[\left(c_p+\frac{1}{2}\right)V^2-V^2\ln\left|V\right|+2AV+B\right]^{1/2}\text{d}V
\end{equation}
Now, it is important to observe that the averaged Lagrangian in Eq.~\eqref{eq: Averaged Lagrangian} depends on six parameters of $\bm{p} = \left[A,B,\omega,k,\beta,\gamma\right]$, but we shall show that these six parameters are not all independent. In fact, it is readily to see that,
\begin{equation}\label{eq: mean average}
    \beta = \overline{V},
\end{equation}
and such relation reduces the six parameters now only to five. Moreover, due to the fixed $2\pi$ period of the periodic function of $\varphi(\theta)$, one can show that,
\begin{equation}\label{eq: non-linear dispersion relation}
    \oint \frac{1}{\sqrt{G(V)}}\text{d}V = \frac{2\pi}{k}.
\end{equation}
The relation in Eq.~\eqref{eq: non-linear dispersion relation} is referred to as the non-linear dispersion relation, and it further reduces the five parameters into four independent ones. Hence, this indicates that we need four modulation equations to form a closed and complete modulation system. To this end, we notice that for the modulated periodic wave, we have to assume all the relevant parameters of $\bm{p}$ as functions of slowly varying spatial and temporal variables. Namely, $\bm{p} = \bm{p}\left(X,T\right)$, where we recall that $X = \delta\xi$ and $T = \delta \tau$. Similar to the method of averaging the conservation laws in Section \ref{sec: Whitham theory}, we introduce the same fast phase as in Section \ref{sec: Whitham theory}, and moreover the generalized fast pseudo-phase $\vartheta = \delta^{-1}\widetilde{S}(X,T)$, where $\widetilde{S}(X,T)$ is a slow pseudo-phase, satisfying,
\begin{equation}\label{eq: pseudo-phase definition}
    \begin{aligned}
    &\vartheta_\xi = \widetilde{S}_X = \beta(X,T),\\
    &\vartheta_\tau = \widetilde{S}_T = -\gamma(X,T),
    \end{aligned}
\end{equation}
and the modulation equations simply follow from the following averaging variational principle,
\begin{equation}
    \delta \int_{-\infty}^{\infty}\int_{-\infty}^{\infty} \mathcal{L}\left(A,B,\omega,k,\beta,\gamma\right)\text{d}X\text{d}T = 0,
\end{equation}
In particular, the variation with respect to $A$ and $B$ yields,
\begin{equation}\label{eq: variations w.r.t to integration const}
    kW_A + \beta = 0, \quad kW_B + \frac{1}{2} = 0.
\end{equation}
which coincide with Eq.~\eqref{eq: mean average} and \eqref{eq: non-linear dispersion relation}, respectively.

Moreover, the variation with respect to the two slow phases of $S$ and $\widetilde{S}$ yields,
\begin{equation}\label{eq: second set of variations}
    \begin{aligned}
        &\left(\mathcal{L}_\omega\right)_T - \left(\mathcal{L}_k\right)_X = 0, \quad k_T + \omega_X = 0. \\
        &\left(\mathcal{L}_\gamma\right)_T - \left(\mathcal{L}_\beta\right)_X = 0 , \quad \beta_T + \gamma_X = 0,
    \end{aligned}
\end{equation}
where we clearly note that the latter two equations in system \eqref{eq: second set of variations} are the two equations of conservation of waves. Finally, the four PDEs in Eq.~\eqref{eq: second set of variations} form the closed and complete Whitham modulation system for the log-KdV equation.

\paragraph{Acknowledgment:} The author would thank Professor Panayotis G.~Kevrekidis from the University of Massachusetts Amherst for fruitful discussions.

\newpage
\bibliographystyle{unsrt}

\bibliography{main}

@article{10.1098/rspa.2013.0462,
    author = {James, Guillaume and Pelinovsky, Dmitry},
    title = {Gaussian solitary waves and compactons in Fermi–Pasta–Ulam lattices with Hertzian potentials},
    journal = {Proceedings of the Royal Society A: Mathematical, Physical and Engineering Sciences},
    volume = {470},
    number = {2165},
    pages = {20130462},
    year = {2014},
    month = {05},
    issn = {1364-5021},
    doi = {10.1098/rspa.2013.0462},
    url = {https://doi.org/10.1098/rspa.2013.0462},
    eprint = {https://royalsocietypublishing.org/rspa/article-pdf/doi/10.1098/rspa.2013.0462/855441/rspa.2013.0462.pdf},
}

@article{EL201611,
title = {Dispersive shock waves and modulation theory},
journal = {Physica D: Nonlinear Phenomena},
volume = {333},
pages = {11-65},
year = {2016},
note = {Dispersive Hydrodynamics},
issn = {0167-2789},
doi = {https://doi.org/10.1016/j.physd.2016.04.006},
url = {https://www.sciencedirect.com/science/article/pii/S0167278916301580},
author = {G.A. El and M.A. Hoefer}
}

@article{10.1063/1.1947120,
    author = {El, G. A.},
    title = {Resolution of a shock in hyperbolic systems modified by weak dispersion},
    journal = {Chaos: An Interdisciplinary Journal of Nonlinear Science},
    volume = {15},
    number = {3},
    pages = {037103},
    year = {2005},
    month = {10},
    issn = {1054-1500},
    doi = {10.1063/1.1947120},
    url = {https://doi.org/10.1063/1.1947120},
    eprint = {https://pubs.aip.org/aip/cha/article-pdf/doi/10.1063/1.1947120/14597729/037103_1_online.pdf},
}

@article{Yang_Biondini_Chong_Kevrekidis_2025, title={A regularized continuum model for travelling waves and dispersive shocks of the granular chain}, volume={1}, DOI={10.1017/S3033426825000038}, journal={Journal of Nonlinear Waves}, author={Yang, Su and Biondini, Gino and Chong, Christopher and Kevrekidis, Panayotis G.}, year={2025}, pages={e2}}

@article{doi:10.1137/S1064827502410633,
author = {Kassam, Aly-Khan and Trefethen, Lloyd N.},
title = {Fourth-Order Time-Stepping for Stiff PDEs},
journal = {SIAM Journal on Scientific Computing},
volume = {26},
number = {4},
pages = {1214-1233},
year = {2005},
doi = {10.1137/S1064827502410633},

URL = { 
    
        https://doi.org/10.1137/S1064827502410633
    
    

},
eprint = { 
    
        https://doi.org/10.1137/S1064827502410633
    
    

}
}

@ARTICLE{Hoefer:2009,
AUTHOR = {Hoefer, M.  and Ablowitz, M. },
TITLE   = {{D}ispersive shock waves},
YEAR    = {2009},
JOURNAL = {Scholarpedia},
VOLUME  = {4},
NUMBER  = {11},
PAGES   = {5562},
DOI     = {10.4249/scholarpedia.5562},
NOTE    = {revision \#137922}
}

@book{whitham2011linear,
  title={Linear and nonlinear waves},
  author={Whitham, Gerald Beresford},
  year={2011},
  publisher={John Wiley \& Sons}
}

@article{CHONG2024103352,
title = {Integrable approximations of dispersive shock waves of the granular chain},
journal = {Wave Motion},
volume = {130},
pages = {103352},
year = {2024},
issn = {0165-2125},
doi = {https://doi.org/10.1016/j.wavemoti.2024.103352},
url = {https://www.sciencedirect.com/science/article/pii/S0165212524000829},
author = {Christopher Chong and Ari Geisler and Panayotis G. Kevrekidis and Gino Biondini}
}

@article{
doi:10.1073/pnas.1517384113,
author = {Steven L. Brunton  and Joshua L. Proctor  and J. Nathan Kutz },
title = {Discovering governing equations from data by sparse identification of nonlinear dynamical systems},
journal = {Proceedings of the National Academy of Sciences},
volume = {113},
number = {15},
pages = {3932-3937},
year = {2016},
doi = {10.1073/pnas.1517384113},
URL = {https://www.pnas.org/doi/abs/10.1073/pnas.1517384113},
eprint = {https://www.pnas.org/doi/pdf/10.1073/pnas.1517384113}}

@misc{chen2019neuralordinarydifferentialequations,
      title={Neural Ordinary Differential Equations}, 
      author={Ricky T. Q. Chen and Yulia Rubanova and Jesse Bettencourt and David Duvenaud},
      year={2019},
      eprint={1806.07366},
      archivePrefix={arXiv},
      primaryClass={cs.LG},
      url={https://arxiv.org/abs/1806.07366}, 
}

@article{PhysRevLett.120.194101,
  title = {Demonstration of Dispersive Rarefaction Shocks in Hollow Elliptical Cylinder Chains},
  author = {Kim, H. and Kim, E. and Chong, C. and Kevrekidis, P. G. and Yang, J.},
  journal = {Phys. Rev. Lett.},
  volume = {120},
  issue = {19},
  pages = {194101},
  numpages = {5},
  year = {2018},
  month = {May},
  publisher = {American Physical Society},
  doi = {10.1103/PhysRevLett.120.194101},
  url = {https://link.aps.org/doi/10.1103/PhysRevLett.120.194101}
}

@article{PhysRevE.75.021304,
  title = {Shock wave structure in a strongly nonlinear lattice with viscous dissipation},
  author = {Herbold, E. B. and Nesterenko, V. F.},
  journal = {Phys. Rev. E},
  volume = {75},
  issue = {2},
  pages = {021304},
  numpages = {8},
  year = {2007},
  month = {Feb},
  publisher = {American Physical Society},
  doi = {10.1103/PhysRevE.75.021304},
  url = {https://link.aps.org/doi/10.1103/PhysRevE.75.021304}
}

@article{PhysRevE.80.056602,
  title = {Stationary shocks in periodic highly nonlinear granular chains},
  author = {Molinari, Alain and Daraio, Chiara},
  journal = {Phys. Rev. E},
  volume = {80},
  issue = {5},
  pages = {056602},
  numpages = {15},
  year = {2009},
  month = {Nov},
  publisher = {American Physical Society},
  doi = {10.1103/PhysRevE.80.056602},
  url = {https://link.aps.org/doi/10.1103/PhysRevE.80.056602}
}

@article{BIONDINI2024134315,
title = {On the Whitham modulation equations for the Toda lattice and the quantitative characterization of its dispersive shocks},
journal = {Physica D: Nonlinear Phenomena},
volume = {469},
pages = {134315},
year = {2024},
issn = {0167-2789},
doi = {https://doi.org/10.1016/j.physd.2024.134315},
url = {https://www.sciencedirect.com/science/article/pii/S0167278924002665},
author = {Gino Biondini and Christopher Chong and Panayotis Kevrekidis}
}

@book{toda2012theory,
  title={Theory of Nonlinear Lattices},
  author={Toda, M.},
  isbn={9783642832192},
  lccn={88021838},
  series={Springer Series in Solid-State Sciences},
  url={https://books.google.com/books?id=ayb7CAAAQBAJ},
  year={2012},
  publisher={Springer Berlin Heidelberg}
}

@article{PhysRevE.90.022905,
  title = {Characterizing traveling-wave collisions in granular chains starting from integrable limits: The case of the Korteweg--de Vries equation and the Toda lattice},
  author = {Shen, Y. and Kevrekidis, P. G. and Sen, S. and Hoffman, A.},
  journal = {Phys. Rev. E},
  volume = {90},
  issue = {2},
  pages = {022905},
  numpages = {12},
  year = {2014},
  month = {Aug},
  publisher = {American Physical Society},
  doi = {10.1103/PhysRevE.90.022905},
  url = {https://link.aps.org/doi/10.1103/PhysRevE.90.022905}
}

@misc{mohapatra2025dambreaksdiscretenonlinear,
      title={Dam breaks in the discrete nonlinear Schr\"odinger equation}, 
      author={Shrohan Mohapatra and Panayotis G. Kevrekidis and Su Yang and Sathyanarayanan Chandramouli},
      year={2025},
      eprint={2507.11529},
      archivePrefix={arXiv},
      primaryClass={nlin.PS},
      url={https://arxiv.org/abs/2507.11529}, 
}

@misc{yang2025dispersiveshockwavesperiodic,
      title={Dispersive shock waves in periodic lattices}, 
      author={Su Yang and Sathyanarayanan Chandramouli and Panayotis G. Kevrekidis},
      year={2025},
      eprint={2511.14549},
      archivePrefix={arXiv},
      primaryClass={nlin.PS},
      url={https://arxiv.org/abs/2511.14549}, 
}

@article{PhysRevA.110.023304,
  title = {Dispersive shock waves in a one-dimensional droplet-bearing environment},
  author = {Chandramouli, Sathyanarayanan and Mistakidis, S. I. and Katsimiga, G. C. and Kevrekidis, P. G.},
  journal = {Phys. Rev. A},
  volume = {110},
  issue = {2},
  pages = {023304},
  numpages = {14},
  year = {2024},
  month = {Aug},
  publisher = {American Physical Society},
  doi = {10.1103/PhysRevA.110.023304},
  url = {https://link.aps.org/doi/10.1103/PhysRevA.110.023304}
}

@article{PhysRevLett.100.084504,
  title = {Piston Dispersive Shock Wave Problem},
  author = {Hoefer, M. A. and Ablowitz, M. J. and Engels, P.},
  journal = {Phys. Rev. Lett.},
  volume = {100},
  issue = {8},
  pages = {084504},
  numpages = {4},
  year = {2008},
  month = {Feb},
  publisher = {American Physical Society},
  doi = {10.1103/PhysRevLett.100.084504},
  url = {https://link.aps.org/doi/10.1103/PhysRevLett.100.084504}
}

@article{https://doi.org/10.1111/sapm.12767,
author = {Sprenger, Patrick and Chong, Christopher and Okyere, Emmanuel and Herrmann, Michael and Kevrekidis, P. G. and Hoefer, Mark A.},
title = {Hydrodynamics of a discrete conservation law},
journal = {Studies in Applied Mathematics},
volume = {153},
number = {4},
pages = {e12767},
doi = {https://doi.org/10.1111/sapm.12767},
url = {https://onlinelibrary.wiley.com/doi/abs/10.1111/sapm.12767},
eprint = {https://onlinelibrary.wiley.com/doi/pdf/10.1111/sapm.12767},
year = {2024}
}

@misc{yang2025quasicontinuumapproximationsnonlineardispersive,
      title={Quasi-continuum approximations for nonlinear dispersive waves in general discrete conservation laws}, 
      author={Su Yang},
      year={2025},
      eprint={2509.04630},
      archivePrefix={arXiv},
      primaryClass={nlin.PS},
      url={https://arxiv.org/abs/2509.04630}, 
}

@article{Abeya_2023,
doi = {10.1088/1751-8121/acb117},
url = {https://doi.org/10.1088/1751-8121/acb117},
year = {2023},
month = {feb},
publisher = {IOP Publishing},
volume = {56},
number = {2},
pages = {025701},
author = {Abeya, Asela and Biondini, Gino and Hoefer, Mark A},
title = {Whitham modulation theory for the defocusing nonlinear Schrödinger equation in two and three spatial dimensions},
journal = {Journal of Physics A: Mathematical and Theoretical}
}

@article{https://doi.org/10.1111/sapm.12651,
author = {Biondini, Gino and Chernyavsky, Alexander},
title = {Whitham modulation theory for the Zakharov–Kuznetsov equation and stability analysis of its periodic traveling wave solutions},
journal = {Studies in Applied Mathematics},
volume = {152},
number = {2},
pages = {596-617},
doi = {https://doi.org/10.1111/sapm.12651},
url = {https://onlinelibrary.wiley.com/doi/abs/10.1111/sapm.12651},
eprint = {https://onlinelibrary.wiley.com/doi/pdf/10.1111/sapm.12651},
year = {2024}
}

@article{Ablowitz_2018,
doi = {10.1088/1751-8121/aabbb3},
url = {https://doi.org/10.1088/1751-8121/aabbb3},
year = {2018},
month = {apr},
publisher = {IOP Publishing},
volume = {51},
number = {21},
pages = {215501},
author = {Ablowitz, Mark J and Biondini, Gino and Rumanov, Igor},
title = {Whitham modulation theory for (2+1)-dimensional equations of Kadomtsev–Petviashvili type},
journal = {Journal of Physics A: Mathematical and Theoretical}
}

@article{PhysRevE.96.032225,
  title = {Whitham modulation theory for the two-dimensional Benjamin-Ono equation},
  author = {Ablowitz, Mark and Biondini, Gino and Wang, Qiao},
  journal = {Phys. Rev. E},
  volume = {96},
  issue = {3},
  pages = {032225},
  numpages = {8},
  year = {2017},
  month = {Sep},
  publisher = {American Physical Society},
  doi = {10.1103/PhysRevE.96.032225},
  url = {https://link.aps.org/doi/10.1103/PhysRevE.96.032225}
}

@misc{yang2025firstordercontinuummodelsnonlinear,
      title={First-order continuum models for nonlinear dispersive waves in the granular crystal lattice}, 
      author={Su Yang and Gino Biondini and Christopher Chong and Panayotis G. Kevrekidis},
      year={2025},
      eprint={2507.07571},
      archivePrefix={arXiv},
      primaryClass={nlin.PS},
      url={https://arxiv.org/abs/2507.07571}, 
}

@book{10.1093/oso/9780192843234.001.0001,
    author = {Carretero-González, Ricardo and Frantzeskakis, Dimitrios J. and Kevrekidis, Panayotis G.},
    title = {Nonlinear Waves \&amp; Hamiltonian Systems: From One To Many Degrees of Freedom, From Discrete To Continuum},
    publisher = {Oxford University Press},
    year = {2024},
    month = {11},
    isbn = {9780192843234},
    doi = {10.1093/oso/9780192843234.001.0001},
    url = {https://doi.org/10.1093/oso/9780192843234.001.0001},
}

@article{doi:10.1137/21M1401243,
author = {Brunton, Steven L. and Budi\v{s}i\'{c}, Marko and Kaiser, Eurika and Kutz, J. Nathan},
title = {Modern Koopman Theory for Dynamical Systems},
journal = {SIAM Review},
volume = {64},
number = {2},
pages = {229-340},
year = {2022},
doi = {10.1137/21M1401243},

URL = { 
    
        https://doi.org/10.1137/21M1401243
    
    

},
eprint = { 
    
        https://doi.org/10.1137/21M1401243
    
    

}
}

@article{KORDA2020599,
title = {Data-driven spectral analysis of the Koopman operator},
journal = {Applied and Computational Harmonic Analysis},
volume = {48},
number = {2},
pages = {599-629},
year = {2020},
issn = {1063-5203},
doi = {https://doi.org/10.1016/j.acha.2018.08.002},
url = {https://www.sciencedirect.com/science/article/pii/S1063520318300988},
author = {Milan Korda and Mihai Putinar and Igor Mezić}
}

\end{document}